\documentclass[a4paper, 11pt]{article}

\usepackage{comment} 
\usepackage[nointegrals]{wasysym}
\usepackage{fullpage} 
\usepackage{graphicx}

\usepackage{cite}
\usepackage{multicol}
\usepackage[perpage]{footmisc} 
\usepackage{sectsty}
\sectionfont{\large}
\usepackage{bm}
\usepackage{float}
\usepackage{amsmath}
\usepackage[utf8]{inputenc}
\usepackage{amssymb}
\usepackage{mathrsfs}
\usepackage{wrapfig}
\usepackage{enumitem}
\usepackage[english]{babel}
\usepackage{comment}
\usepackage{bbold}  
\usepackage[usenames,dvipsnames]{xcolor} 
\usepackage{xcolor,hyperref}
\usepackage{setspace}
\usepackage{abstract}
\usepackage{tablefootnote}

\usepackage[T1]{fontenc}
\usepackage[margin=0.94in]{geometry}

\def\Mpl{M_{\rm P}}

\usepackage{enumitem}

\begin{document}
\null\hfill IPMU25-0012 \\
\null\hfill YITP-25-36 \\
\vspace*{\fill}
\begin{center}
    \Large\textbf{\textsc{ \textcolor{Black}{On the cosmological degrees of freedom of Proca field with non-minimal coupling to gravity } }}

    \normalsize\textsc{Antonio De Felice$^{1}$, Anamaria Hell$^{2}$}
\end{center}

\begin{center}
$^{1}$ \textit{Center for Gravitational Physics and Quantum Information,\\
Yukawa Institute for Theoretical Physics,\\ Kyoto University,\\ 606-8502, Kyoto, Japan}\\
    $^{2}$ \textit{Kavli IPMU (WPI), UTIAS,\\ The University of Tokyo,\\ Kashiwa, Chiba 277-8583, Japan}
\end{center}
\thispagestyle{empty} 

\renewcommand{\abstractname}{\textsc{\textcolor{Black}{Abstract}}}

\begin{abstract}
 We study Proca theory with non-minimal coupling to gravity through the Ricci tensor and Ricci scalar interactions. We show that in the homogeneous and isotropic Universe together with cosmological constant, the temporal component of the vector field acquires a background value. As a result, we show that the theory propagates an additional degree of freedom, with respect to the generalized Proca theories, whose kinetic term suggests the presence of  {several} strong coupling regimes that depend on the value of the background solution, the combination {and vanishing} of coupling constants, {together with a scale-dependent one.}
   We show in addition, that the speed of propagation for this mode vanishes, indicating the presence of a {another} type of strong coupling. To further investigate this, we extend our analysis to the Bianchi Type I Universe, with the most general solution for the vector field. We show that the extra {degree of freedom} remains in the theory. Among the modes, we further show that the mode with vanishing speed of propagation is still present, pointing to the strong coupling. In addition, we discover a mode with scale-dependent strong coupling (vanishing kinetic term), 
  one mode that propagates only in one single direction and two unstable modes. 
\end{abstract}
 
\vfill
\small
\href{mailto:antonio.defelice@yukawa.kyoto-u.ac.jp}{\text{antonio.defelice@yukawa.kyoto-u.ac.jp}}\\
\href{mailto:anamaria.hell@ipmu.jp}{\text{anamaria.hell@ipmu.jp}}
\vspace*{\fill}

\clearpage
\pagenumbering{arabic} 
\newpage

\section{Introduction }

Models based on vector fields are among the most natural candidates for solving the puzzles underlying the evolution and structure of our Universe. Already appearing in the standard model of particle physics, they have recently attracted considerable attention as dark matter candidates. Moreover, besides the scalar fields, they have also been explored as a way to account for the early- and late-time acceleration of the universe, and have been studied in connection with magnetogenesis and anisotropy.

To get interesting cosmological phenomena, however, one needs to depart from standard electrodynamics, as this theory is conformally invariant. In order to achieve this, many models consider a non-trivial kinetic coupling for the vector field of the form $f(\phi)F_{\mu\nu}F^{\mu\nu}$, with $\phi$ being an inflaton, or some other field \cite{Demozzi:2009fu, Watanabe:2009ct, Ratra:1991bn, Bamba:2003av, Martin:2007ue, Emami:2009vd, 2011PhRvD..84l3525C, Fujita:2012rb, Ferreira:2013sqa, Kobayashi:2014sga, Ferreira:2014mwa, Domenech:2015zzi, Campanelli:2015jfa, Fujita:2016qab, Tasinato:2014fia}. They preserve the gauge invariance, while the conformal invariance is explicitly broken. Another interesting possibility is to break gauge invariance, by considering the massive vector theory known as the Proca theory. This is famously known as the theory of a dark photon \cite{Essig:2013lka, Fabbrichesi:2020wbt}, a dark matter candidate. Besides the coupling with the visible matter through non-gravitational interactions, it allows for the dark matter production via the gravitational particle production thus making a compelling candidate which is at the same time completely dark \cite{Graham:2015rva, Ahmed:2020fhc, Kolb:2020fwh}. More generally, one can also add to the vector fields a potential $V(A_{\mu}A^{\mu})$ \cite{Koivisto:2008xf}, derivative couplings \cite{  Mukohyama:2016npi, Tasinato:2014eka,  Heisenberg:2014rta, Allys:2015sht, BeltranJimenez:2016rff, Allys:2016jaq, Heisenberg:2016eld, Kimura:2016rzw, Emami:2016ldl} and pseudo-scalar couplings \cite{Anber:2009ua}, or consider non-Abelian generalizations of the vector fields \cite{ Adshead:2012kp, Maleknejad:2011jw, Maleknejad:2011sq}. Such models have very interesting phenomenology, ranging from the non-Gaussianity, parity-violating gravitational waves, baryogenesis, magnetogenesis, anisotropy, black holes {(including primordial ones)}, %\ADF{black holes,}
and the possibility to underline inflation and dark energy  \cite{Barnaby:2010vf, Barnaby:2011vw, Barnaby:2011qe, Crowder:2012ik,  Namba:2015gja, Maleknejad:2016qjz, Obata:2016tmo, Obata:2016oym, Caprini:2014mja, Fujita:2015iga, Adshead:2016iae, Anber:2015yca, Cado:2016kdp, Linde:2012bt, McDonough:2016xvu, Pajer:2013fsa, Adshead:2012qe, Sheikh-Jabbari:2012tom, Maleknejad:2011jr, Dimastrogiovanni:2012ew, Adshead:2013nka, Namba:2013kia, Nieto:2016gnp, Adshead:2016omu, DeFelice:2016uil, DeFelice:2016yws, Watanabe:2010fh, Gumrukcuoglu:2010yc, Emami:2010rm, Emami:2011yi, Soda:2012zm, Shiraishi:2012xt, Bartolo:2012sd, Emami:2013bk, Heisenberg:2017xda, DeFelice:2024bdq, Shiraishi:2013vja, Lyth:2013sha, Biagetti:2013qqa, Ohashi:2013qba, Abolhasani:2013bpa, Naruko:2014bxa, Abolhasani:2015cve, Heisenberg:2016wtr, Garcia-Serna:2025dhk, Garnica:2021fuu, Murata:2021vnb, GallegoCadavid:2020dho, Gorji:2020vnh, Gorji:2019ttx, Firouzjahi:2018wlp, Heisenberg:2018mxx, BeltranJimenez:2017cbn, Martinez:2024gsj, Savvidy:2022ies, Gomez:2020sfz, Alexander:2014uza, BeltranAlmeida:2014aau }.

 {Among the possible vector theories, an especially interesting case is the non-minimal coupling between the vector field and gravity. With maximally two time-derivatives, such coupling can appear with a Ricci scalar, Ricci tensor, or through an Einstein tensor, which is a specific combination of the previous two.\footnote{Another possibility for the non-minimal coupling involves a coupling with the Gauss-Bonnet term, which we will not consider in this work.} Recently, these models have gained significant attention in the context of enhanced dark matter production via the gravitational particle production in comparison to its minimal version \cite{Cembranos:2023qph, Ozsoy:2023gnl, Capanelli:2024rlk}. They naturally yield dynamical solutions for the vector field, driving the expansion of the Universe \cite{Koivisto:2008xf}, and show promise in generating the primordial magnetic fields, driving inflation, or even as candidates of dark energy \cite{Turner:1987bw, Golovnev:2008cf, Kanno:2008gn, Dimopoulos:2008rf, Turner:1987bw, Demozzi:2009fu}.  }

 A common approach to studying these theories is by treating the vector field as a spectator one in a cosmological background. This is under the assumption that the influence of the vector field on a given background is negligible. In this case, however, not all values of the coupling constants characterizing the interactions are allowed. As pointed out in \cite{Himmetoglu:2008zp, Himmetoglu:2009qi}, the non-minimal interactions can give rise to ghost instabilities, characterized by the wrong sign of the kinetic term for the scalar mode, which is absent in the massless, minimally coupled theory. The non-minimal coupling can also give rise to the runaway modes, whose gradient terms have a wrong kinetic sign, or super-luminal propagation \cite{Grzadkowski:2024oaf, Capanelli:2024rlk, Capanelli:2024pzd}. Notably, the aforementioned instabilities could be avoided if the general non-minimal coupling is replaced with the combination of the Ricci scalar and Ricci tensor that equals the Einstein tensor \cite{Ozsoy:2023gnl}. However, due to the presence of the Ricci tensor, the theory can also introduce the strong coupling of the tensor modes in flat space-time, which can be resolved if one defines it in the disformal frame \cite{Hell:2024xbv}.

The previous picture can become greatly modified if one considers a vector field with a non-vanishing background value. In general, the background values for the spatial components of the vector fields easily introduce an anisotropy. In the FLRW Universe, to avoid it, one can consider a large number of randomly oriented independent non-interacting fields, or particular combinations such as vector triplets \cite{Golovnev:2008cf, Golovnev:2009ks}. Alternatively, one can study a single non-minimally coupled vector field, albeit {only} in the anisotropic universe, such as Bianchi Type I \cite{Himmetoglu:2008zp}.

Due to the non-vanishing background value of the spatial component of the vector field, the metric perturbations mix with the vector field, and cannot be neglected any longer. As pointed out in \cite{Himmetoglu:2008zp, Himmetoglu:2009qi}, in the case of the coupling with the Ricci scalar, where a single spatial component of the vector field or spatial part of the triplet of vector fields acquires a non-vanishing, time-dependent expectation value, the ghost instability can be again present in the theory and associated with the longitudinal component of the vector field. Moreover, since the metric and vector field components do not necessarily decouple, this can lead to tachyonic instability for the gravitational waves for large fields vector inflation \cite{Golovnev:2008hv} (see also \cite{Golovnev:2009ks} for the further study on cosmological perturbations for the small field models).

Notably, in addition to the possible appearance of instabilities, in \cite{Golovnev:2009rm, Golovnev:2011yc} it was noticed that if one considers $N$ vector fields with non-minimal coupling to the Ricci scalar, the model might propagate an additional degree of freedom in a background that is not homogeneous. In particular, one combination of temporal components of all of the $N$ vector fields would become dynamical due to the coupling with the Ricci scalar, which contains second-order time derivatives of one of the gravitational potentials. Thus one might lose a constraint that is otherwise present in the Proca theory and propagate an additional mode. 
This is contrasted to the generalized Proca scenario, where, by construction, the authors were building an action for $A_\mu$, such that its temporal element $A_0$ was not propagating on any background. This construction is similar to the construction of general Horndeski in the context of scalar-tensor theories {\cite{Horndeski:1974wa, Nicolis:2008in, Deffayet:2011gz, Deffayet:2009wt, Kobayashi:2011nu}}\footnote{Or the construction of the Lovelock scalars in the context of pure gravity.} and was aimed to leave the theory only with three propagating degrees of freedom (besides the gravitational ones).

In this work, we will follow along the lines of  {the} idea {given in \cite{Golovnev:2009rm, Golovnev:2011yc}}, and explore if an extra degree of freedom appears in a single Proca theory with a quadratic, non-minimal coupling to gravity. In particular, we will show that  in the presence of a cosmological constant and a homogeneous and isotropic universe, this precisely happens. {In particular, we will show that this happens even in a homogeneous and isotropic Universe. }Surprisingly, however, we will show that while there exist no-ghost conditions for which the scalar is well-behaved, its speed of propagation is zero. Since this usually implies a strong coupling, we will extend our analysis also to the Bianchi Type I background, and study the theory for non-vanishing spatial and temporal background values of the vector field.

We will show that in Bianchi I, {one of the modes with vanishing speed of propagation still remains in the theory. In addition, } 

at least in the parameter space close to the homogeneous and isotropic Friedmann–Lemaître–Robertson–Walker (FLRW) limit, {we will find that a} new scale-dependent \footnote{{By this we mean a kinetic term proportional to $(k^2-\mathcal{A}q^2)^2$, with $\mathcal{A}$ a positive numerical value.}} strong coupling arises together with new Laplacian instabilities (i.e.\ the speed of propagation $c_s$ having an imaginary part for at least two of the propagating modes) will occur for the even-mode perturbations. This scenario suggests  
a problematic phenomenology for these theories, as the homogeneous and isotropic limit {for the linearized theory} seems either {unstable or difficult to treat due to non-linearities}
(at least in the absence of matter fields, therefore this analysis would hold for inflationary and/or late-time dynamics), and one needs to fix all these issues somehow if one believes these theories have something to do with our world. {For example, the apparent issues might be resolved by taking into account non-linear terms and verifying the behavior of the theory at a background that is less isotropic.}

The paper is structured as follows. In Sec.~II we study the homogeneous and isotropic realization of the theory. In particular, we will study the scalar, vector, and tensor decomposition for the perturbation fields, and look for no-ghost and stability conditions for the spacetime to be stable. In Sec.~III, we focus our attention on the background solutions that could have interesting phenomenological applications and study their stability both regarding different initial conditions on the background and against the linear perturbation propagation. Sec.~IV is devoted to the analysis of the same theory on the homogeneous but anisotropic Bianchi-I manifold, to try to understand the unclear points present in the FLRW perturbation analysis. In Sec.~V we report our stability analysis results and explain the phenomenology of the perturbation stability on the anisotropic manifold. Finally, our conclusions are presented in Sec.~VI.

\section{The homogeneous and isotropic universe}
The goal of this paper is to count the number of degrees of freedom (dof) in a single Proca theory with non-minimal coupling to gravity. This theory is described by the following action: 
\begin{equation}\label{NMProcaAction}
    \begin{split}
        S=\int d^4x\sqrt{-g}&\left[\frac{\Mpl^2}{2}\left(R-2\Lambda\right)-\frac{1}{4}F_{\mu\nu}F^{\mu\nu}-\frac{m^2}{2}A_{\mu}A^{\mu}-\frac{1}{2}\left(\beta_1  RA_{\mu}A^{\mu}+\beta_2 R^{\mu\nu}A_{\mu}A_{\nu}\right)\right],
    \end{split}
\end{equation}
where
\begin{equation}
    F_{\mu\nu}=\nabla_{\mu}A_{\nu}-\nabla_{\nu}A_{\mu}
\end{equation}
is the field strength tensor, $m$ is the mass of the vector field, and $\beta_{1,2}$ are the coupling constants that characterize the non-minimal coupling of the vector field to gravity. Here, we have also added the cosmological constant. In the following, we will consider the theory with it, and comment about the case when it is set to zero at the end of this section. 

As a first step, we will study this theory in the homogeneous and isotropic universe, given by the  Friedmann–Lemaître–Robertson–Walker (FLRW) metric: 
\begin{equation}
    ds^2=-N^2(t)dt^2+a^2(t)dx^idx^j.
\end{equation}
Here, $a(t)$ is the scale factor, and $N(t)$ is the lapse. This metric describes a homogeneous and isotropic universe and determines the possible background values for the vector field. In particular, due to the symmetries, we can notice that for a single vector field, only the temporal component can have a non-vanishing background value. In contrast, the spatial value must vanish, unless one considers a triplet of vector fields or N randomly oriented ones.  Since we are only interested in a case with a single vector field, we will choose its background value as: 
\begin{equation}
    A_{\mu}=\left(N(t)A_0(t), 0, 0, 0 \right). 
\end{equation}
As a next step, let us consider the background equations of motion. For this, we vary the action with respect to the lapse, the scale factor, and the temporal component. By varying the action with respect to $N(t)$, and solving for the cosmological constant $\Lambda$, and subsequently setting $N(t)=1$ everywhere, we find the first Friedman equation: 
\begin{equation}\label{Feq1}
    \Lambda=\frac{1}{2 \Mpl^{2}}\left[\left(\left(6 \beta_{1}+6 \beta_{2}\right) H^{2}+m^{2}\right) A_{0}^{2}+12\dot{A}_0 \left(\beta_{1}+\frac{\beta_{2}}{2}\right) A_{0} H +6 H^{2} \Mpl^{2}\right]. 
\end{equation}
Here, the dot denotes a derivative with respect to the coordinate time $t$. 
By varying the action with respect to the scale factor, we find: 
\begin{equation}\label{acceqOrgigi}
\begin{split}
    &\left(\left(2 \beta_{2}+2 \beta_{1}\right) A_{0}^{2}+2 \Mpl^{2}\right) \frac{\ddot{a}}{a}
   +\ddot{A}_0 \left(2 \beta_{1}+\beta_{2}\right) A_{0} +\left(H^{2} \beta_{2}+H^{2} \beta_{1}+\frac{1}{2} m^{2}\right) A_{0}^{2}\\&+4 H \dot{A}_0 \left(\beta_{1}+\beta_{2}\right) A_{0}+2 \dot{A}_0^{2} \beta_{1}+\dot{A}_0^{2} \beta_{2}+\Mpl^{2} \left(H^{2}-\Lambda \right)=0. 
    \end{split}
\end{equation}
By substituting the above expression for $\Lambda$ in it, we find the acceleration equation: 
\begin{equation}
   \begin{split}
   \left(\left( \beta_{1}+ \beta_{2}\right) A_{0}^{2}+ \Mpl^{2}\right) \frac{\Ddot{a}}{a}
   = &-\left(\beta_{1}+\frac{\beta_{2}}{2}\right) A_{0} \Ddot{A}_0+H^{2} \left(\beta_{1}+\beta_{2}\right) A_{0}^{2}\\&+H \dot{A}_0 \left(\beta_{1}-\frac{\beta_{2}}{2}\right) A_{0}+H^{2} \Mpl^{2}-\dot{A}_0^{2} \beta_{1}-\frac{\dot{A}_0^{2} \beta_{2}}{2} 
   \end{split}
\end{equation}
Finally, by varying with respect to the temporal component of the vector field, we find: 
\begin{equation}
    \left(6 \beta_{1} a \,H^{2}+a \,m^{2}+3\Ddot{a}(2  \beta_{1}+ \beta_{2})\right) A_{0}=0
\end{equation}
This equation yields two branches for the $A_0$ component---when it takes a vanishing and a non-vanishing value. In this work, we will be interested in the case $A_0\neq 0$. Then, we can solve the previous equation for the second derivative in the scale factor:
\begin{equation}\label{tempeq}
    \Ddot{a} =
-\frac{a \left(6 H^{2} \beta_{1}+m^{2}\right)}{6 \beta_{1}+3 \beta_{2}}
\end{equation}
Altogether, the background equations of motion with $N(t)=1$ which we will use to find the behavior of the perturbations are then given by: 
\begin{equation}\label{FLRWbeom}
    \begin{split}
        \Lambda=&\frac{1}{2 \Mpl^{2}}\left[\left(\left(6 \beta_{1}+6 \beta_{2}\right) H^{2}+m^{2}\right) A_{0}^{2}+12\dot{A}_0 \left(\beta_{1}+\frac{\beta_{2}}{2}\right) A_{0} H +6 H^{2} \Mpl^{2}\right],\\\\
\Ddot{a} =& 
-\frac{a \left(6 H^{2} \beta_{1}+m^{2}\right)}{6 \beta_{1}+3 \beta_{2}}\,,\\\\
\Ddot{A}_0 =& 
\frac{2 \left(\beta_{1}+\beta_{2}\right) A_{0} \left(\left(12 \beta_{1}+3 \beta_{2}\right) H^{2}+m^{2}\right)}{3 \left(2 \beta_{1}+\beta_{2}\right)^{2}}+\frac{H \left(2 \beta_{1}-\beta_{2}\right) \dot{A}_0}{2 \beta_{1}+\beta_{2}}-\frac{\dot{A}_0^2}{A_0}\\&+\frac{2 \Mpl^{2} \left(m^{2} a^{2}+3 \left(4 \beta_{1}+\beta_{2}\right) a^{2} H^{2}\right)}{3 a^{2}A_{0} \left(2 \beta_{1}+\beta_{2}\right)^{2}}\,.
    \end{split}
\end{equation}

We can notice that the second equation yields an exact solution for the scale factor. In particular, we can obtain an accelerating universe in the following two cases: 
\begin{equation}
    \begin{split}
        &1) \qquad 6 H^{2} \beta_{1}+m^{2}>0\qquad \text{and}\qquad 2\beta_1+\beta_2<0\qquad \text{or}\\
        &2) \qquad 6 H^{2} \beta_{1}+m^{2}<0\qquad \text{and}\qquad 2\beta_1+\beta_2>0.
    \end{split}
\end{equation}
However, we should note that the above equations hold only for $\beta_2\neq-2\beta_1$. The special case for which $\beta_2=-2\beta_1$ should be treated separately, and we will not assume it in this work, but rather assume the general relation among the coupling constants.\footnote{This case corresponds to the generalized Proca case, for which only three degrees of freedom propagate in addition to the gravity gravitational waves.} 

We should note that the above equations yield interesting solutions. While in this section we will use them in the form (\ref{FLRWbeom}) to study the perturbations, substituting them to the corresponding action, the subsequent section will study the background solutions, and focus on the islands of stability, combining both the stability of the background solutions and also taking into account finding for the perturbations that we will uncover in this section.

Let us now perturb the metric, and the vector field around their background values: 
\begin{equation}
    g_{\mu\nu}= g_{\mu\nu}^{(0)}+\delta  g_{\mu\nu},\qquad \qquad A_{\mu}=A_{\mu}^{(0)}+\delta  A_{\mu}. 
\end{equation}
Here, $g_{\mu\nu}^{(0)}$ and $A_{\mu}^{(0)}$ are the background fields that we have previously specified, and which satisfy the above background equations of motion (\ref{FLRWbeom}). The perturbations will be decomposed according to the group of spatial rotations. For the metric, this is given by: 
\begin{equation}
    \begin{split}
        \delta g_{00}&=-2\phi\\
        \delta g_{0i}&=a(t)\left(S_i+B_{,i}\right)\\
        \delta g_{ij}&=a^2(t)\left(2\psi \delta_{ij}+2E_{,ij}+F_{i,j}+F_{j,i}+h_{ij}^T\right),
    \end{split}
\end{equation}
where the comma denotes the derivative with respect to the spatial component. The vector field  perturbations will be given by:  
\begin{equation}
    \delta A_{0}=A\qquad\text{and}\qquad \delta A_i=a(t)A_i^T+\chi_{,i}. 
\end{equation}

The vector modes are divergent-less: 
\begin{equation}
    S_{i,i}=0,\qquad F_{i,i}=0,\qquad \text{and}\qquad A_{i,i}^T=0. 
\end{equation}
In addition to being divergent-less, the tensor modes are also traceless: 
\begin{equation}
    h_{ij,j}^T=0\qquad \text{and}\qquad h_{ii}^T=0.
\end{equation}
At linearized order, the scalar, vector, and tensor modes decouple. Thus, we will treat them separately, starting from the scalar perturbations. 

\subsection{Scalar perturbations} 
In this subsection, we will consider only the scalar perturbations. They are characterized by six fields: $\phi, B, E, \psi, A$ and $\chi$. Under infinitesimal coordinate transformations, 
\begin{equation}\label{infcoo}
    x^{\mu}\to\Tilde{x}^{\mu}=x^{\mu}+\xi^{\mu}, 
\end{equation}
with 
\begin{equation}
    \xi^{\mu}=(\xi^0, \xi^i),\qquad \xi^{Ti}_{,i}=0,\qquad \xi^i\equiv \delta_{ij} \left( \xi_{j}^T+\zeta_{,j}\right),
\end{equation}
the scalar potentials transform as: 
\begin{equation}
    \begin{split}
        &\phi\to\Tilde{\phi}=\phi-\dot{\xi}^0\qquad \qquad B\to \Tilde{B}=B+\frac{1}{a}\xi^0-a\dot{\zeta} \qquad\qquad \psi\to \tilde{\psi}=\psi-\frac{\dot{a}}{a}\xi^0\\
        &E\to \tilde{E}=E-\zeta \qquad\qquad A\to\tilde{A}=A-\left(A_0\xi^0\right)^.\qquad\qquad \;\; \chi\to\tilde{\chi}=\chi-A_0\xi^0. 
    \end{split}
\end{equation}
Using them, we can form the gauge-invariant potentials: 
\begin{equation}
  \begin{split}
        &\Phi=\phi+\left[a\left(B-a\dot{E}\right)\right]^.,\qquad\Psi=\psi+\dot{a}\left(B-a\dot{E}\right)\\ & \underline{A}=A+\left[A_0a\left(B-a\dot{E}\right)\right]^. ,\qquad\text{and}\qquad\underline{\chi}= \chi+A_0\left(B-a\dot{E}\right).
  \end{split}
\end{equation}
By choosing the conformal gauge: 
\begin{equation}
    E=0\qquad\text{and}\qquad  B=0,
\end{equation}
the scalar potentials match with the gauge-invariant ones, among which the first two correspond to the Bardeen potentials \cite{Mukhanov:2005sc, Bardeen:1980kt}. Therefore, in the remainder of this subsection, we will work in this gauge without the loss of generality. 
By expanding the action (\ref{NMProcaAction}) to the quadratic order in perturbations, and performing several integrations by parts, we find the following Lagrangian density: 
\begin{equation}\label{FRWStartingL}
    \begin{split}
        \mathcal{L}&=c_1(t)\dot{\psi}^{2}+c_2(t)\dot{\phi} \dot{\psi}+c_3(t)\dot{\psi} \dot{A}+c_4(t)\dot{\chi}\Delta\dot{\chi}\\
        &+c_5(t)\dot{\phi} A+c_6(t)\dot{\phi} \psi+c_7(t)\dot{\psi} \phi+c_8(t)\dot{\psi} A +c_9(t) \dot{\psi}\Delta \chi+c_{10}(t)A\Delta\dot{\chi}\\
        &+c_{11}(t)\psi\Delta \psi +c_{12}(t)A \Delta A+c_{13}(t)\chi\Delta \chi +c_{14}(t)\phi\Delta \phi \\
        & +c_{15}(t)\phi\Delta \chi +c_{16}(t)A\Delta \psi+c_{17}(t)\phi\Delta \psi+ c_{18}(t)A\Delta \phi+ c_{19}(t)\phi\Delta \chi\\&+c_{20}(t)\phi\psi+c_{21}(t)\psi A +c_{22}(t)\phi A+c_{23}(t)\phi^2+c_{24}(t)\psi^2+c_{25}(t)A^2
    \end{split}
\end{equation}
Here, $c_1 ... c_{25}$ are time-dependent coefficients that depend on the Planck mass, the mass of the vector field, the scale factor, and the background value of the temporal component of the vector field. Their detailed expressions are given in the Appendix \ref{APP1.1}. By expressing the fields in terms of the Fourier modes,
\begin{equation}
    X(t,\Vec{x})=\int \frac{d^3k}{\left(2\pi\right)^{\frac{3}{2}}}X_k(t)e^{i\Vec{k}\Vec{x}}, \qquad\text{where}\qquad X=\left\{\phi,\psi,A,\chi\right\}
\end{equation}
and by substituting the background equations of motion (\ref{FLRWbeom}), we further find: 
\begin{equation}\label{FRWF1}
    \begin{split}
        S=&\int dtd^3k\left[ d_1\dot{\psi}_k\dot{\psi}_{-k}+d_2\left(\dot{\phi}_{k} \dot{\psi}_{-k}+\dot{\phi}_{-k}\dot{\psi}_{k}\right)+d_3\left(\dot{\psi}_k\dot{A}_{-k}+\dot{\psi}_{-k}\dot{A}_{k}\right)+d_4\dot{\chi}_k\dot{\chi}_{-k}\right.\\&\left.+d_5\left(\dot{\phi}_{k} A_{-k}+\dot{\phi}_{-k} A_{k}\right)+d_6\left(\dot{\phi}_{k}\psi_{-k}+\dot{\phi}_{-k}\psi_{k}\right)+d_7\left(\dot{\psi}_{k} \phi_{-k}+\dot{\psi}_{-k} \phi_{k}\right)\right.\\&\left. +d_8\left(\dot{\psi}_{k} A_{-k}+\dot{\psi}_{-k} A_{k}\right) +d_9 \left(\dot{\psi}_k\chi_{-k}+\dot{\psi}_{-k}\chi_{k}\right)+d_{10}\left(A_{k}\dot{\chi}_{-k}+A_{-k}\dot{\chi}_{k}\right)\right.\\&\left. +d_{11}\psi_k\psi_{-k} +d_{12}A_k A_{-k}+d_{13}\chi_k \chi_{-k}+d_{14}\phi_k\phi_{-k}+
        d_{15}\left(\phi_k\chi_{-k}+\phi_{-k}\chi_{k}\right) \right.\\&\left.+d_{16}\left(\psi_kA_{-k}+\psi_{-k}A_{k}\right)+d_{17}\left(\phi_{k} \psi_{-k}+\phi_{-k} \psi_{k}\right)+ d_{18}\left(A_{k}\phi_{-k}+A_{-k}\phi_{k}\right)\right].
    \end{split}
\end{equation}

Here, the coefficients $d_1 ... d_{18}$ depend on the time and momenta. They are explicitly given in the appendix (\ref{APP1.2}). {On the one hand, we can notice that there are no terms such as $\dot{\phi}^2$ or $\dot{A}^2$. On the other, such terms can still contribute to the kinetic matrix due to the couplings corresponding to the coefficients $d_2$ and $d_3$.}  
By performing the substitution, 
\begin{equation}
    \phi_k=\phi_{2k}+\frac{A_{k}}{A_0(t)},
\end{equation}
  followed by a partial integration of $A\dot{A}$, we can notice that the term with the $d_3$ coefficient cancels, with  $d_2$ still remaining in the action. {Therefore, we can notice that $A$ is not propagating. }
By varying the  action with respect to $A$, we find its corresponding constraint. We then solve it and substitute it back to the action. This makes the resulting action a function of only three fields, $\phi_2, \psi$, and $\chi$. We further partially integrate the terms of the form $\phi_2\dot{\phi}_2$, $\psi\dot{\psi}$ and $\chi\dot{\chi}$. While in the resulting action, all terms will appear to come with time derivatives, the determinant of the corresponding kinetic matrix will vanish, indicating that they are dependent. By performing the following substitutions: 
\begin{equation}
    \psi_k=\psi_{2k}-\frac{HA_0}{\dot{A}_0}\phi_{2k}\qquad\text{and}\qquad \chi_k=\chi_{2k}-\frac{A_0^2}{\dot{A}_0}\phi_{2k},
\end{equation}
followed by partially integrating $\psi_2\dot{\psi}_2$, we find that the resulting kinetic matrix has a non-propagating term in $\phi_2$. By finding its constraint, solving it, and substituting it back to the action, we find an expression that only depends on $\psi_2$ and $\chi_2$. We further partially integrate terms of the form $\psi_k\dot{\psi}_k$ and $\chi_k\dot{\chi}_k$. Then, by further substituting 
\begin{equation}
   \chi_2= \frac{1}{k}\chi_3+\frac{2 \left(\left(\beta_{1}+\beta_{2}\right) A_{0}^{2}+\Mpl^{2}\right) a A_{0} }{2 \left(\beta_{1}+\beta_{2}\right) a H A_{0}^{2}+2 \dot{A}_0 A_{0} \left(\beta_{1}+\frac{\beta_{2}}{2}\right) a +2 a H \,\Mpl^{2}}\psi_2,
\end{equation}
we find: 
\begin{equation}\label{FRWF2}
    \begin{split}
        S=\int dtd^3k&\left[ e_1(t,k)\dot{\psi}_{2k}\dot{\psi}_{2,-k}+e_2(t)\dot{\chi}_{3k}\dot{\chi}_{3,-k}+e_3(t)\left(\dot{\chi}_{3k}\psi_{2,-k}+\dot{\chi}_{3,-k}\psi_{2,k}\right)\right.\\&\left. +e_4(t)\psi_{2k}\psi_{2,-k}+e_5(t)\chi_{3k}\chi_{3,-k}+e_6(t)\left(\chi_{3k}\psi_{2,-k}+\chi_{3,-k}\psi_{2,k}\right)\right],
    \end{split}
\end{equation}
where $e_1 ... e_6$ are time and momenta-dependent coefficients, with the last four given in the Appendix. The first two coefficients give us the no-ghost conditions: 
\begin{equation}
    \begin{split}
        e_1=\frac{3 \dot{A}_0^{2} A_{0}^{2} a^{5} \left(\beta_{1}+\frac{\beta_{2}}{2}\right)^{2} \left(\left(\beta_{1}+\beta_{2}\right) A_{0}^{2}+\Mpl^{2}\right)}{\left(\left(\beta_{1}+\beta_{2}\right) a H A_{0}^{2}+\dot{A}_0 A_{0} \left(\beta_{1}+\frac{\beta_{2}}{2}\right) a +a H \,\Mpl^{2}\right)^{2}}\qquad \text{and}\qquad e_2(t)=\frac{a}{2}
    \end{split}
\end{equation}
Clearly, for 
 \begin{equation}
     \left(\beta_{1}+\beta_{2}\right) A_{0}^{2}+\Mpl^{2}>0 \qquad\to \qquad  \left(\beta_{1}+\beta_{2}\right)>0
 \end{equation}
the two modes have a positive sign in front of the kinetic term and are thus well-behaved. However, we can also notice that $ e_1$ vanishes in %two
{three} different limits. The first limit is when the background $A_0(t)$ vanishes. The second is for $\beta_{1}{\to}-\frac{\beta_{2}}{2}$. The third appears when $\beta_1\to 0$ and $\beta_2\to 0$.  {At the same time, the theories with $A_0=0$, $\beta_{1}=-\frac{\beta_{2}}{2}$ or $\beta_1=\beta_2= 0$ have only one scalar mode.\footnote{{Note that the  $\beta_{1}=-\frac{\beta_{2}}{2}$ corresponds to the special case of Generalized Proca theory which includes only the coupling with the Einstein tensor, and $\beta_1=\beta_2= 0$  is the Proca theory with minimal coupling to gravity.} } Intuitively, we can expect that this apparent discontinuity in the dof at linearized order will lead to a strong coupling regime, with the presence of a scale at which the non-linear terms become dominant. In fact, this was shown to appear for interacting Proca theory,  Yang-Mills theory with mass added by hand, massive and interacting Kalb-Ramond and three-form theories as well as in theories of modified gravity \cite{Vainshtein:1972sx, Dvali:2006su, Deffayet:2001uk, Gruzinov:2001hp, Dvali:2007kt, Chamseddine:2010ub, Alberte:2010it, Mukohyama:2010xz, deRham:2010ik, deRham:2010kj, Chamseddine:2012gh, Tasinato:2014eka, Chamseddine:2018gqh, Hell:2021wzm, Hell:2023mph, Heisenberg:2014rta, Heisenberg:2020xak,  BeltranJimenez:2016rff, Hell:2021oea}. So it would be natural to expect that the above additional mode could enter a strong coupling regime. }

Let us now find the speed of propagation for the modes. For this, we find the equations of motion for the two scalars, $\psi_2$ and $\chi_3$. Then, we solve the equation for  $\psi_2$ in terms of the first derivative of  $\chi_3$:
\begin{equation}
    \dot{\chi}_3=\Psi_E(\Ddot{\psi}_2, \dot{\psi}_2, \psi_2, \chi_3)
\end{equation}
We substitute this solution back to the equation of motion for $\chi_3$, eliminating with the above equation $\dot{\chi}_3$ and $\Ddot{\chi}_3$. The resulting equation can be solved for $\chi_3$: 
\begin{equation}
    \chi_3=X_E(\Ddot{\psi}_2, \dot{\psi}_2, \psi_2).
\end{equation}
Substituting this back to the equation of motion for $\psi_2$, we find a closed equation for $\psi_2$, that includes terms with fourth-order time derivatives. By assuming a plane wave solution:
\begin{equation}
    \psi_{2k}=A_k e^{i\omega t},
\end{equation}
This equation becomes in the limit of high momenta: 
\begin{equation}
    \omega^4 B_4(t) + \omega^3 B_3(t) + k^2\left( \omega^2 B_2(t) + \omega B_1(t) + B_0(t)\right)=0.
\end{equation}
Here, $B_1 ... B_0$ are time-dependent coefficients. By assuming the dispersion relation of the form
\begin{equation}
    \omega=\lambda k^p,\qquad p\geq1,
\end{equation}
we find a non-trivial solution for $p=1$, 
\begin{equation}
    \lambda^2\left(\lambda^2B_4+B_2\right)\sim0.
\end{equation}
This leads to the following values for the speed of propagation: 
\begin{equation}
    c_s^2=0,
\end{equation}
and 
\begin{equation}
    c_s^2=\frac{3\Mpl^2+A_0^2\left(2\beta_2^2+3\beta_1+4\beta_2\right)}{3\left(\Mpl^2+A_0^2(\beta_1+\beta_2)\right)}. 
\end{equation}
We can notice that in the absence of the Ricci tensor coupling ($\beta_2=0$), the speed of propagation is unity. This means that in contrast to the Ricci scalar coupling, the Ricci tensor one distorts the speed of propagation. 

Overall, in this subsection, we have found two scalar degrees of freedom in the Proca theory with non-minimal coupling to gravity. This is in contrast to the minimal theory, which only describes a longitudinal mode, a single scalar degree of freedom. The reason for its appearance is twofold. It arises due to the non-minimal coupling of the vector field with the Ricci scalar and Ricci tensor, and because the background value of the temporal component of the vector field is not vanishing. One should note that in this example, we have sourced this component by using the cosmological constant. However, we could have also considered self-interactions, such as in generalized Proca theory \cite{DeFelice:2016yws}. As long as $A_0^{(0)}\neq 0$, one can expect that the extra dof arises in the theory.
Nevertheless, here we have shown that even though this extra dof appears, its speed of propagation vanishes. This indicates that the higher-order terms might be important, or in other words, a breakdown in the perturbation theory. To explore this further, in the next section, we will study the theory in a Bianchi Type I Universe. Before that, we will study the remaining degrees of freedom - the vector and tensor modes.

\subsection{Vector perturbations}

Let us first consider the vector modes. Under the infinitesimal coordinate transformations, (\ref{infcoo}), the metric vector potentials transform as: 
\begin{equation}
    S_i\to\tilde{S}_i=S_i-a\dot{\xi}_i\qquad\text{and}\qquad F_i\to\tilde{F}_i=F_i-\xi^T_i,
\end{equation}
While the transverse modes of the vector field, $A_i^T$ are gauge-invariant. The gauge invariant combination of the metric vector modes is: 
\begin{equation}
    V_i=S_i-a\dot{F}_i. 
\end{equation}
In this subsection, we will work on the gauge 
\begin{equation}
    F_i=0,
\end{equation}
in which $S_i$ coincides with the gauge-invariant variable.

By expanding the action (\ref{NMProcaAction}) to the quadratic order in perturbations, and by integrating by parts, we find: 
\begin{equation}\label{FRWF3}
    \begin{split}
        \mathcal{L}=&v_1(t)\dot{A}_i^T\dot{A}_i^T+v_2(t)A_i^T\Delta A_i^T+v_3(t)A_i^TA_i^T+
        v_4(t)S_i\Delta S_i+v_5(t)S_iS_i+v_6(t)A_i^T\Delta S_i
    \end{split}
\end{equation}
Here, $v_1 ... v_6$ are time-dependent coefficients, given in the Appendix \ref{APP1.4}. We can write the vector modes in the Fourier space as
\begin{equation}
    A^T_i=\sum_{\sigma=1,2}\int \frac{d^3k}{(2\pi)^{\frac{3}{2}}}\varepsilon_i^{\sigma} v_k^{\sigma}(t)e^{i\Vec{k}\Vec{x}},\qquad  S_i=\sum_{\sigma=1,2}\int \frac{d^3k}{(2\pi)^{\frac{3}{2}}}\varepsilon_i^{\sigma} s_k^{\sigma}(t)e^{i\Vec{k}\Vec{x}}
\end{equation}

By substituting the background equations of motion into the previous action, we find: 
\begin{equation}
  \begin{split}
        S=\sum_{\sigma=1,2} \int dt d^3k & \left[u_1\dot{v}_k^{\sigma}\dot{v}_{-k}^{\sigma}+u_2v_k^{\sigma}v_{-k}^{\sigma}+u_3s_k^{\sigma}s_{-k}^{\sigma}+u_4\left(v_k^{\sigma}s_{-k}^{\sigma}+s_k^{\sigma}v_{-k}^{\sigma}\right)\right]. 
  \end{split}
\end{equation}
Here we have used the normalization of the polarization vectors. The coefficients in the action are given by:
\begin{equation}
    \begin{split}
        &u_1(t,k)=\frac{a^{3}}{2}\\
        &u_2(t,k)=-\frac{a}{2 \beta_{1}+\beta_{2}}\left[\frac{m^{2} \left(\beta_{2}-\frac{1}{2}\right) a^{2}}{3}+4 \beta_{2} \left(\frac{\beta_{2}}{4}+\frac{1}{8}+\beta_{1}\right) a^{2} H^{2}+\left(\beta_{1}+\frac{\beta_{2}}{2}\right) k^{2}\right] \\
        &u_3(t,k)=\frac{a}{4}\left[\left(\beta_1+\beta_2\right)k^2A_0^2+\Mpl k^2\right]\\
        &u_4(t,k)=\frac{\beta_{2}  k^{2} A_{0} a}{4}
    \end{split}
\end{equation}
We can notice that $s_k^{\sigma}$ is not propagating. By varying the action with respect to it, solving the constraint, and substituting it back to the action, we find the action written in terms of the two propagating modes: 
\begin{equation}
    \begin{split}
          S=\sum_{\sigma=1,2} \int dt d^3k \left[\frac{a^{3}}{2}\dot{v}_k^{\sigma}\dot{v}_{-k}^{\sigma}+\Tilde{u}(t,k) v_k^{\sigma}v_{-k}^{\sigma},
\right]
    \end{split}
\end{equation}
where 
\begin{equation}
   \begin{split}
        \Tilde{u}=&-\frac{a}{2 \left(\left(\beta_{1}+\beta_{2}\right) A_{0}^{2}+\Mpl^{2}\right) \left(\beta_{1}+\frac{\beta_{2}}{2}\right)}\left[\frac{1}{3} \left(A_{0}^{2}\left(\beta_{1}+\beta_{2}\right) m^{2} \left(\beta_{2}-\frac{1}{2}\right) a^{2}\right) \right. \\& \left. +A_{0}^{2}\left(4 \beta_{2} \left(\beta_{1}+\beta_{2}\right) \left(\frac{\beta_{2}}{4}+\frac{1}{8}+\beta_{1}\right) a^{2} H^{2}+\left(\frac{1}{2} \beta_{2}^{2}+\beta_{1}+\beta_{2}\right) \left(\beta_{1}+\frac{\beta_{2}}{2}\right) k^{2}\right) \right. \\ &\left. +\left(\frac{m^{2} \left(\beta_{2}-\frac{1}{2}\right) a^{2}}{3}+4 \beta_{2} \left(\frac{\beta_{2}}{4}+\frac{1}{8}+\beta_{1}\right) a^{2} H^{2}+\left(\beta_{1}+\frac{\beta_{2}}{2}\right) k^{2}\right) \Mpl^{2}\right]
   \end{split}
\end{equation}
We can notice that the no-ghost condition for the vector modes is always satisfied. 
By varying the action with respect to $v_k^{\sigma}$, we find the corresponding equation of motion from which we can read off the speed of propagation. In particular, by assuming a solution in the form of a plane wave, 
\begin{equation}
    v_k^{\sigma}=A_k^{\sigma}e^{ikx},
\end{equation}
we find that the speed of propagation is given by:
\begin{equation}
   c_V^2= \frac{A_{0}^{2} \beta_{2}^{2}+2 A_{0}^{2} \beta_{1}+2 A_{0}^{2} \beta_{2}+2 \Mpl^{2}}{2 A_{0}^{2} \beta_{1}+2 A_{0}^{2} \beta_{2}+2 \Mpl^{2}}
\end{equation}
Similarly to the case of scalar modes, the speed of sounds becomes unity in the absence of the interactions with the Ricci tensor. 
Therefore, in addition to the two scalar modes, this theory describes two vector modes, for which the no-ghost condition is always satisfied.

\subsection{Tensor perturbations} 
Finally, let us study the tensor modes. After expanding the initial action to second order in perturbations and integrating by parts, we find that the Lagrangian density corresponding to the (gauge-invariant) tensor dof is given by: 
\begin{equation}
    \mathcal{L}=w_1(t)\dot{h}_{ij}^T\dot{h}_{ij}^T+w_2(t)h_{ij}^T\Delta h_{ij}^T+ w_3(t)h_{ij}^Th_{ij}^T,
\end{equation}
where
\begin{equation}
    \begin{split}
        w_1(t)=&\frac{ a^{3}}{8}\left[\left(\beta_{1}+\beta_{2}\right) A_{0}^{2}+\Mpl^{2}\right]\\
        w_2(t)=&\frac{a}{8}\left(A_{0}^{2} \beta_{1}+\Mpl^{2}\right) \\
        w_3(t)=&-\frac{a}{8}\left[\left(m^{2} A_{0}^{2}+4 \left(\beta_{1}+\frac{\beta_{2}}{2}\right) \left( \Ddot{A}_0 A_{0}+\dot{A}_0^{2}\right)-2 \Lambda \,\Mpl^{2}\right) a^{2}+8 a^2 H\dot{A}_0 \left(\beta_{1}+\beta_{2}\right) A_{0}\right.\\&\left.+\frac{2a^2}{3} \left(-6 H^{2}+R \right) \left( \left(\beta_{1}+\beta_{2}\right)A_{0}^{2}+\Mpl^{2}\right) +2 \left(\left(\beta_{1}+\beta_{2}\right) A_{0}^{2}+\Mpl^{2}\right) a^{2} H^{2}\right]
    \end{split}
\end{equation}
Here, $R$ is the background value of the Ricci scalar.  
It is convenient to express the tensor modes in the Fourier space: 
\begin{equation}
    h^T_{ij}=\sum_{\sigma=1,2}\int \frac{d^3k}{(2\pi)^{\frac{3}{2}}}\varepsilon_i^{T\sigma} h_k^{\sigma}(t)e^{i\Vec{k}\Vec{x}},
\end{equation}
where $\varepsilon_i^{T\sigma} $ are the corresponding polarization vectors. Then, by substituting the background equations of motion into the action, we find: 

\begin{equation}
    \begin{split}
          S=\sum_{\sigma=1,2} \int dt d^3k \left[\frac{\left(\left(\beta_{1}+\beta_{2}\right) A_{0}^{2}+\Mpl^{2}\right) a^{3}}{8}\dot{h}_k^{\sigma}\dot{h}_{-k}^{\sigma}-\frac{a \,k^{2} \left(A_{0}^{2} \beta_{1}+\Mpl^{2}\right)}{8} h_k^{\sigma}h_{-k}^{\sigma},
\right]
    \end{split}
\end{equation}
where we have normalized the polarization vectors. The no-ghost condition is given by:
\begin{equation}
    \left(\beta_{1}+\beta_{2}\right) A_{0}^{2}+\Mpl^{2}>0, 
\end{equation}
and coincides with the no-ghost condition for the scalar modes. 
By finding the equation of motion and assuming the plane wave solution, we find that the speed of propagation is given by:
\begin{equation}
    c_T^2=\frac{A_{0}^{2} \beta_{1}+\Mpl^{2}}{A_{0}^{2} \beta_{1}+A_{0}^{2} \beta_{2}+\Mpl^{2}}
\end{equation}
Similarly to the remaining modes, the Ricci tensor coupling causes the change in the speed of propagation, while in the case of only Ricci scalar coupling, it equals unity. {The non-trivial speed of propagation was also found in \cite{Golovnev:2008hv} and generalized to other p-forms in \cite{Kobayashi:2009hj}. }

\subsection{Comment on the cosmological constant}

In the previous section, we studied the perturbations of the Proca theory, with non-minimal coupling to gravity, in the FLRW background, by expanding the action up to second order in the perturbations, and by substituting the equations for the background, given in (\ref{FLRWbeom}), into the action. The background equations of motion were solved for $\Lambda\neq 0$, therefore, one might wonder if it is possible to get another result by setting $\Lambda=0$. In this case, instead of using the set of background equations (\ref{FLRWbeom}), it is more convenient to solve the equation (\ref{Feq1}) for $\dot{A}_0$, and substitute this, together with (\ref{tempeq}), an equation for $\ddot{a}$, into (\ref{acceqOrgigi}), solving it for $\ddot{A}_0$. Then, the set of the background equations of motion is given by the equations for $\{\dot{A}_0, \ddot{A}_0, \ddot{a}\}$. As a result, the substitutions for the scalar modes will be slightly different than as described in the previous section. However, one can easily check that the conditions for the absence of instability will remain the same in the two cases, for all types of modes and independent of the value of the cosmological constant. 

Overall, in this section, we have found that the Proca theory with non-minimal coupling to gravity perturbed around the most general background in a homogeneous and isotropic universe propagates six modes---two scalar, two vector, and two tensor modes. The reason why this appears can be traced to the presence of the non-vanishing background value of the temporal component of the vector field, $A_0(t)$, and the general non-minimal coupling in the Lagrangian with gravity. This can be also traced to the no-ghost conditions for the scalar modes---while one remains in the theory always, the kinetic term of the other vanishes when $A_0(t)=0$, (or when $\beta_2+2\beta_1\approx0$\footnote{The exact value $\beta_1=-\beta_2/2$ corresponds to a generalized Proca case, which, having less degrees of freedom, is not continuously obtained from the Lagrangian we have considered here.}), which also indicates strong coupling when $A_0(t)\to 0$. 

While the no-ghost condition for the vector modes is always satisfied, the no-ghost conditions for the tensor and scalar modes are equivalent and allow for the presence of ghosts if violated. %\ADFcomm{Some tiny changes.}\hell{Thanks! I agree. } 
Furthermore, we can notice that if the coupling with the Ricci tensor vanishes, $\beta_2=0$, vector, tensor, and one of two the scalar dofs have a speed of propagation that becomes unity. Otherwise, we obtain a non-trivial form similar to the no-ghost conditions in all modes, allowing for the presence of instabilities. 

The remaining scalar mode has a vanishing speed of propagation. This usually indicates that the non-linear terms become dominant, thus rendering our analysis non-trustable. To gain more insight into this behavior, we will explore the theory on the Bianchi Type I background in the latter sections. Before that, we will complete the analysis of the homogeneous and isotropic background by studying the background solutions, which when combined with the conditions in this section will give us the stability conditions for the theory. 

\section{Background solutions and the islands of stability}

In the previous section, we have found regimes that might constrain the parameters of the theory to guarantee the absence of ghosts or instabilities. In this section, we will investigate this further, by considering the background solutions for the homogeneous and isotropic universe and combining them with the previous results.

The background equations of motion we have found in the previous section were obtained by varying the action with respect to the lapse, the scale factor, and the $A_0$ component of the vector field, and then setting $N(t)=1$. In particular, we have found that the constraint equation was given by: 
\begin{equation}\label{constrainteq}
 \begin{split}
        6\left(\left( \beta_{1}+ \beta_{2}\right) A_{0}^{2}+ \Mpl^{2}\right)H^{2} +12 \dot{A}_0 \left(\beta_{1}+\frac{\beta_{2}}{2}\right) A_{0}H+ \left(A_{0}^{2} m^{2}-2 \Lambda \,\Mpl^{2}\right)=0
 \end{split},
\end{equation}
while the variation with respect to the scale factor and the temporal component of the vector field has yielded respectively 
\begin{equation}\label{acceqFLRW2}
\begin{split}
 &3 \left(\left(\beta_{2}+\beta_{1}\right) A_{0}^{2}+\Mpl^{2}\right)H^{2}+4 A_{0} \dot{A}_0 \left(\beta_{2}+\beta_{1}\right)H+2 \left(\left(\beta_{2}+\beta_{1}\right) A_{0}^{2}+\Mpl^{2}\right) \dot{H}\\&+ A_{0} \left(\beta_{2}+2 \beta_{1}\right) \ddot{A}_0+\frac{m^2}{2} A_{0}^{2} + \dot{A}_0^{2} \beta_{2}- \Lambda \,\Mpl^{2}+2 \dot{A}_0^{2} \beta_{1}=0
, 
\end{split}
\end{equation}
and
\begin{equation}\label{tempFLRW}
   [3 \left(\beta_{2}+2 \beta_{1}\right) \dot{H}+3 \left(4 \beta_{1}+\beta_{2}\right)H^{2}+m^{2}] A_{0}=0. 
\end{equation}

While in the previous section, we have used the combination of these equations to study the perturbations, in this section, we will focus on their solutions. Our goal is to determine the solution for the Hubble parameter $H(t)$ and the temporal component $A_0(t)$. Thus, among the above three equations, we can notice that it would be best to focus on the constraint equation, and the one for the temporal field. In particular, by assuming $A_0(t)\neq 0$, we find the equation for the Hubble parameter: 
\begin{equation}\label{dotH}
   \dot{H} = 
-\frac{12 \beta_{1} H^{2}+3 H^{2} \beta_{2}+m^{2}}{6 \beta_{1}+3 \beta_{2}}
\end{equation}
In turn, the constraint equation can be used to determine the temporal component: 
\begin{equation}\label{dottemp}
    \dot{A}_0=\frac{-3\left(\left( \beta_{1}+\beta_{2}\right) H^{2}-m^{2}\right) A_{0}^{2}+\Mpl^{2} \left(-3 H^{2}+\Lambda \right)}{3\left(2 \beta_{1}+ \beta_{2}\right) H A_{0}}, 
\end{equation}
or, to put it more simply, it's square: 
\begin{equation}\label{dottempsq}
    \dot{Y}=\frac{-6\left(\left( \beta_{1}+\beta_{2}\right) H^{2}-m^{2}\right) Y+2\Mpl^{2} \left(-3 H^{2}+\Lambda \right)}{3\left(2 \beta_{1}+ \beta_{2}\right) H }, 
\end{equation}
In the following, we will focus on solving these two equations for large times, assuming that $\beta_1\neq-\frac{\beta_2}{2}$ when both of them are different from zero. One can easily check that the remaining equation (\ref{acceqFLRW2}) will be automatically satisfied.

By solving now solve Eq.\ \eqref{dotH}, we notice the existence of two different dynamics for the variable $H$. If $4\beta_{1}+\beta_{2}>0$, then we have
\begin{equation}
H=-\frac{\sqrt{3}m}{3\sqrt{4\beta_{1}+\beta_{2}}}\,\tan\!\left(\frac{\sqrt{12\beta_{1}+3\beta_{2}}m(t-t_{0})}{6\beta_{1}+3\beta_{2}}\right),
\end{equation}
which leads to a singularity of spacetime ($R\to\infty$) in a finite time. We will discard this case and consider, from now on, instead the case of $4\beta_{1}+\beta_{2}<0$. Then we have
\begin{equation}
H=\frac{m\,\tanh\!\left(-\frac{\sqrt{-12\beta_{1}-3\beta_{2}}\,m(t-t_{0})}{6\beta_{1}+3\beta_{2}}\right)}{\sqrt{-12\beta_{1}-3\beta_{2}}}\,,
\end{equation}
so that for $t\gg t_{0}$, we have
\begin{equation}
H\to\pm\frac{m}{\sqrt{-12\beta_{1}-3\beta_{2}}}\,.
\end{equation}
To have an expanding solution at late times,\footnote{We will not consider the phenomenology of the universe for $H<0$,
leading to an exponentially small universe, $a=a_{0}e^{-|H|t}$.} $t\gg t_{0}$, we require
\[
2\beta_{1}+\beta_{2}<0\,.
\]
This last condition, combined with the other condition $4\beta_{1}+\beta_{2}<0$,
leads to the following sets of solutions:
\begin{equation}
\{\beta_{1}<0,\beta_{2}<-2\,\beta_{1}\}\,,\qquad{\rm or}\qquad\{\beta_{2}<-4\beta_{1},\beta_{1}\geq0\}\,.
\end{equation}
Unless otherwise stated, from now on, we will work considering these constraints to hold in any case.
On defining $mt=\tau$ (considering $m>0$), $\lambda=\Lambda/m^{2}$,
and $\bar{Y}=Y/M_{P}^{2}$, Eq.\ \eqref{dottempsq} can be rewritten as
\begin{equation}
    \bar{Y}' \left(2\beta_{1}+\beta_{2}\right) \sqrt{-12 \beta_{1}-3 \beta_{2}}\, \tanh \gamma(\tau)=
    8 \left(\frac{\bar{Y}}{2}-\lambda \right)\! \left(\beta_{1}+\frac{\beta_{2}}{4}\right)
    -2[1+\left(\beta_{1}+ \beta_{2}\right) \bar{Y}] \tanh^{2}\gamma(\tau)\,,
    \label{eq:barY_prime}
\end{equation}
where
\begin{equation}
    \gamma(\tau)=-\frac{\sqrt{-12 \beta_{1}-3 \beta_{2}}\, \left(\tau -\tau_0  \right)}{6 \beta_{1}+3 \beta_{2}}\,,
\end{equation}

and a prime denotes differentiation with respect to $\tau$. Let us now try to find approximate asymptotic solutions.

\subsection{Standard fixed points}\label{sec:stand_fix_points}
Let us assume that $t\gg t_{0}$, ($\tau\gg \tau_0$), and approximate the dynamical equation, Eq.\ (\ref{eq:barY_prime}), as
\begin{align}
-\sqrt{-12\beta_{1}-3\beta_{2}}\,\bar{Y}'\,(2\beta_{1}+\beta_{2}) & =-\left(2\beta_{1}-\beta_{2}\right)\bar{Y}+2+2\lambda\left(4\beta_{1}+\beta_{2}\right),
\end{align}
which can be integrated exactly.The case  $\beta_{2}>2\beta_{1}$, will be discussed later in Sec.\ \ref{sec:ext_fix_points}, as in this case, the dynamics are dominated by a growing solution {while in the appendix we will perform a more general fixed point analysis and analyze the stability of the solutions}. In this section, we will consider instead the case 
$2\beta_{1}>\beta_{2}$,  for which %\ADF{for which}
the homogeneous solution will be suppressed. We will call here by standard fixed points, the solutions for which asymptotically $\dot{Y}\to0$, and $H\to H_{{\rm dS}}$, where $H_{{\rm dS}}\equiv\frac{m}{\sqrt{-12\beta_{1}-3\beta_{2}}}>0$.

Since for $t\gg t_{0}$, ${\rm sech}^{2}\gamma(\tau)\to0$, exponentially, we find that these fixed points exist provided that
\begin{equation}
2+2\lambda\left(4\beta_{1}+\beta_{2}\right)-\left(2\beta_{1}-\beta_{2}\right)\bar{Y}\approx0\,,\qquad\bar{Y}\geq0,\quad2\beta_{1}>\beta_{2}\,.
\end{equation}
This implies
\begin{equation}
\bar{Y}=\frac{2+2\lambda\left(4\beta_{1}+\beta_{2}\right)}{2\beta_{1}-\beta_{2}}\geq0\,.\label{eq:Y_fin}
\end{equation}
or the following three sets of existing solutions: 
\begin{equation}
\left\{ \lambda<-\frac{1}{4\beta_{1}+\beta_{2}},\beta_{1}<-\frac{\beta_{2}}{4},\beta_{2}<0,\frac{\beta_{2}}{2}<\beta_{1}\right\} ,
\end{equation}
so that standard fixed points do exist for the above set in parameters space. If these late points are approachable, that depends on the property $Y\geq0$, during the whole dynamics. Notice that the existence of standard fixed points does not imply the solution will approach them in any case.
There are some choices of initial conditions, which go through negative $Y$'s as a transient before reaching, finally, the fixed points.
In this case, we will have that the fixed point is not reached, as our analysis stops when $Y\to0$.  What happens when $Y$ crosses zero?
In this case, we will have that $A_{0}^{2}\to0^{+}$, and this means that one will have the constraints coming from this particular limit.
This case has been discussed when we have studied the behavior of the perturbations dynamics for the scalar modes. 

What happens instead when $2\beta_{1}-\beta_{2}>0$, allowing for $\dot{Y}\to0$, but $\frac{1+\lambda\,(4\beta_{1}+\beta_{2})}{2\beta_{1}-\beta_{2}}<0$, not allowing for positive or zero solutions for $Y$? Numerically, we find that $Y$ crosses 0 going to a negative-$Y$ standard fixed point. Also, in this case, we will have to deal with the $A_{0}^2\to0^{+}$ limit.

\subsection{Extended fixed points}\label{sec:ext_fix_points}

As already noticed in Sec.\ \ref{sec:stand_fix_points},  different dynamics occur if $\beta_2\geq 2\beta_1$, while keeping as already stated above, before, here, and after, the conditions $2\beta_1+\beta_2<0$ and $4\beta_1+\beta_2<0$. So we want to discuss here this possibility in parameter space. 

For this goal, let us reconsider the dynamical equation in the late-time regime then

\begin{equation}
-\sqrt{-12\beta_{1}-3\beta_{2}}\,\bar{Y}'\,(2\beta_{1}+\beta_{2})=\left(\beta_{2}-2\beta_{1}\right)\bar{Y}+2+2\lambda\left(4\beta_{1}+\beta_{2}\right),\label{eq:exp_gro}
\end{equation}
where we have further assumed that $2\beta_{1}+\beta_{2}<0$ and $4\beta_{1}+\beta_{2}<0$, again to have an expanding universe in the far future. Here we study the case of solutions which at late time have $\bar{Y}'\neq0$. This happens in the following two cases.

\subsubsection{Special case 1: Linear growth of $A_{0}^{2}$ }

Special case, {Let us consider} $\beta_{2}=2\beta_{1}$. In this case, the above equation
simplifies to
\begin{equation}
6\sqrt{-2\beta_{1}}\,\bar{Y}'\,(-\beta_{1})=1+6\lambda\beta_{1},
\end{equation}
and the solution makes sense if and only if $1+6\lambda\beta_{1}>0$
(i.e.\ if not, $Y$ would cross 0), or
\begin{equation}
\left\{ \beta_{2}=2\beta_{1},\lambda<-\frac{1}{6\beta_{1}},\beta_{1}<0\right\} .
\end{equation}

\subsubsection{Special case 2: Exponential growth for $A_{0}^{2}$ }

The
solution of Eq.\ (\ref{eq:exp_gro}) reads %\hell{Correct. }
\begin{align}
\frac{Y}{2M_{P}^{2}}&=c_{1}\exp\!\left[-\frac{(\beta_{2}-2\beta_{1})\,mt}{\sqrt{-12\beta_{1}-3\beta_{2}}(2\beta_{1}+\beta_{2})}\right]+\frac{1+\lambda(4\beta_{1}+\beta_{2})}{2\beta_{1}-\beta_{2}}\nonumber\\
&\approx c_{1}\exp\!\left[-\frac{(\beta_{2}-2\beta_{1})\,mt}{\sqrt{-12\beta_{1}-3\beta_{2}}(2\beta_{1}+\beta_{2})}\right].
\end{align}

\begin{figure}[ht]
    \centering
    \includegraphics[width=0.47\linewidth]{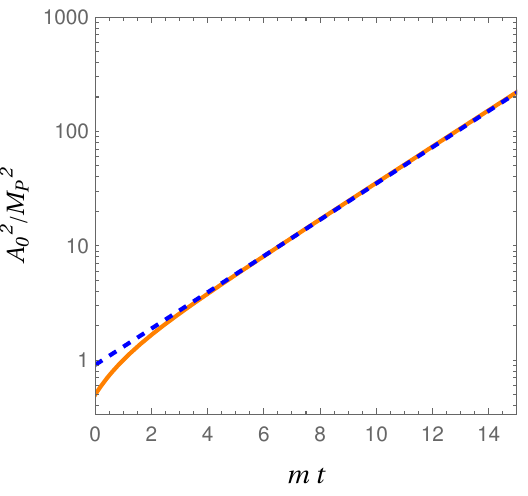}\hfill
    \includegraphics[width=0.47\linewidth]{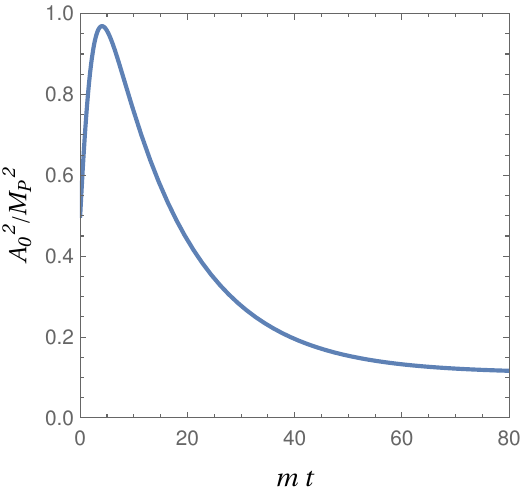}
    \caption{Numerical solutions of the differential equation, Eq.~\eqref{eq:barY_prime}, for ${\bar Y}\equiv A_0^2/\Mpl^2$. In the left panel, we show exponential growth as a possible solution for the dynamics. This plot was obtained by setting the following parameters: $\beta_1=-3$, $\beta_2=2$, $\Lambda/m^2=1/20$, and $\tau_0=-1$; and initial condition $\bar{Y}(0)=1/2$. The dashed line represents its analytical approximation, $\bar{Y}=\frac{91}{100}e^{2mt/\sqrt{30}}$. In the figure on the right, we instead show the solution for a parameter space element inside the set $\mathcal{S}_1$, more precisely for $\beta_1=-1$, $\beta_2=-4$, and $\Lambda/m^2=1/9$. We further set $\tau_0=-1$ and $\bar{Y}(0)=1/2$. This second case corresponds to a fixed point solution which is stable under all the background and perturbation conditions.}
    \label{fig:exp_&_S1}
\end{figure}

This exponential-growth case corresponds to the following parameter space:
\begin{equation}
\left\{ \beta_{2}\le0,\beta_{1}<\frac{\beta_{2}}{2}\right\} ,\left\{ 0<\beta_{2},\beta_{1}<-\frac{\beta_{2}}{2}\right\} .
\end{equation}
Note that for this case, a choice of initial conditions is necessary, because, for instance, if $c_{1}<0$, the solution will develop to $Y\to-\infty$. Nonetheless, there will be plenty of initial conditions for which the solution grows to positive $Y$'s. An example of these exponentially growing solutions is given in Fig.~\ref{fig:exp_&_S1}.

\subsection{Stability conditions from perturbation analysis}
We need to check whether the perturbation fields are stable or not on the background solutions found above. As for the linearly or exponentially growing background solutions for $\bar{Y}$, one finds that the tensor modes set the extra constraints
\begin{equation}
    (\beta_1+\beta_2)\bar{Y}+1\approx (\beta_1+\beta_2)\bar{Y} >0\,,\quad
    c_T^2=\frac{1+\beta_1\bar{Y}}{1+(\beta_1+\beta_2)\bar{Y}}\approx
    \frac{\beta_1}{\beta_1+\beta_2}>0\,.
\end{equation}

\begin{figure}[ht]
    \centering
    \includegraphics[width=0.47\linewidth]{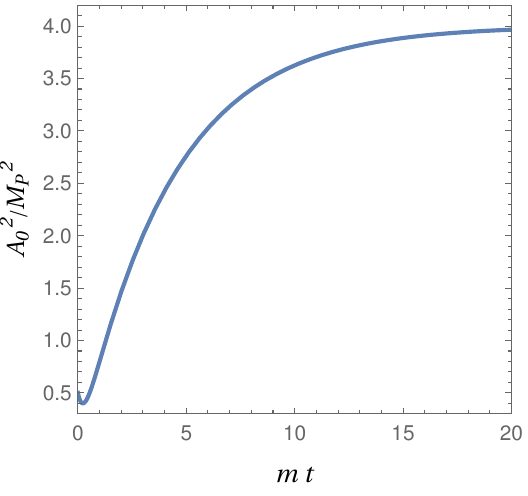}\hfill
    \includegraphics[width=0.47\linewidth]{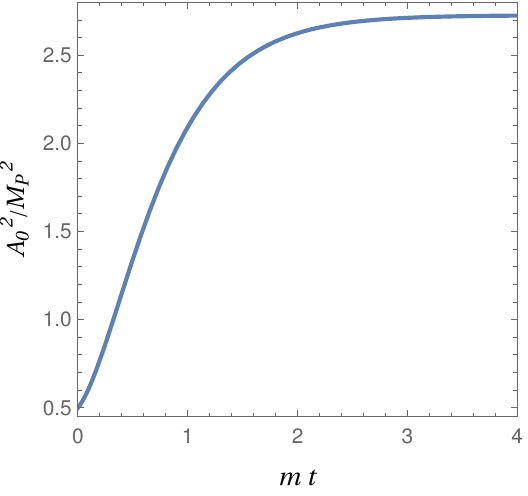}
    \caption{Numerical solutions of the differential equation, Eq.~\eqref{eq:barY_prime}, for ${\bar Y}\equiv A_0^2/\Mpl^2$. In the left panel, we show a solution belonging to the allowed set of parameter space $\mathcal{S}_2$. This plot was obtained by setting the following parameters: $\beta_1=-\frac{1}{30}$, $\beta_2=-\frac{1}{10}$, $\Lambda/m^2=4$, and $\tau_0=-1$; and initial condition $\bar{Y}(0)=1/2$. In the figure on the right, we instead show the solution for a parameter space element inside the set $\mathcal{S}_3$, more precisely for $\beta_1=\frac{1}{30}$, $\beta_2=-\frac{3}{10}$, and $\Lambda/m^2=3$. We further set $\tau_0=-1$ and $\bar{Y}(0)=1/2$. The fixed point solutions shown in these two figures, since they are determined by parameters belonging to the sets defined in Eq.~\eqref{eq:S123}, are stable under all the background and perturbation conditions.}
    \label{fig:S2_&_S3}
\end{figure}

None of the linearly/exponentially growing solutions are compatible with these last constraints, and as such, have to be discarded. If we now consider the solutions for the standard fixed points, first we define
\begin{equation}
    {\bar Y}_{\rm fin}\equiv \frac{2+2\lambda\left(4\beta_{1}+\beta_{2}\right)}{2\beta_{1}-\beta_{2}}\,.
\end{equation}
In this case, in addition to the background constraints, we need to impose that the no-ghost conditions for scalars, tensor, and speed of propagation for all the modes are positive, namely 

\begin{alignat}{6}
    4\beta_1+\beta_2&<0\,,&\qquad&\textrm{no future singularities;}\label{eq:cstr1}\\
    2\beta_1+\beta_2&<0\,,&& \textrm{future expanding solution;}\label{eq:cstr2}\\
    \beta_2&<2\beta_1\,, &&\textrm{removing solutions unstable under tensor modes;}\label{eq:cstr3}\\
    {\bar Y}_{\rm fin}&>0\,,&& \textrm{reality condition on $A_0^2$;}\\
    (\beta_1+\beta_2){\bar Y}_{\rm fin}+1&>0\,,&& \textrm{no ghost-condition;}\\
    \frac{1+{\bar Y}_{\rm fin}\beta_1}{1+{\bar Y}_{\rm fin}(\beta_1+\beta_2)}&>0\,,&&
    \textrm{positive speed of propagation for tensor modes;}\\
    \frac{2+{\bar Y}_{\rm fin}(\beta_2^2+2\beta_1+2\beta_2)}{2[1+{\bar Y}_{\rm fin}(\beta_1+\beta_2)]}&>0\,,&&
    \textrm{positive speed of propagation for vector modes;}\\
    \frac{3+{\bar Y}_{\rm fin}(2\beta_2^2+3\beta_1+4\beta_2)}{ 3[1+{\bar Y}_{\rm fin}(\beta_1+\beta_2) ]}&>0\,,&&
    \textrm{positive speed of propagation for scalar modes.}
\end{alignat}
leading to any one of the following sets of allowed parameter space 

\begin{align}
\mathcal{S}_1&\equiv\left\{\beta_{2}\le -\frac{1}{2}, 
\lambda <-\frac{1}{4 \beta_{1}+\beta_{2}}, 
\beta_{1}<-\frac{\beta_{2}}{4}, \frac{\beta_{2}}{2}<\beta_{1}, 
-\frac{1}{2 \beta_{2}+2 \beta_{1}}<\lambda\right\},\\
\mathcal{S}_2&\equiv \left\{
-\frac{1}{2}<\beta_{2}, \lambda <-\frac{1}{4 \beta_{1}+\beta_{2}}, 
\beta_{1}<0, \beta_{2}<0, \frac{\beta_{2}}{2}<\beta_{1}, 
\frac{-4 \beta_{2}^{2}-12 \beta_{1}-5 \beta_{2}}{2 \left(2 \beta_{2}^{2}+3 \beta_{1}+4 \beta_{2}\right) \left(4 \beta_{1}+\beta_{2}\right)}<\lambda
\right\},\\
%&\left\{\beta_{1} = 0, -\frac{1}{2}<\beta_{2}, \lambda <-\frac{1}{\beta_{2}}, \beta_{2}<0, \frac{-4 \beta_{2}-5}{4 \beta_{2} \left(\beta_{2}+2\right)}<\lambda\right\},\\
\mathcal{S}_3&\equiv\left\{0\leq \beta_{1}, -\frac{1}{2}<\beta_{2}, 
\lambda <-\frac{1}{4 \beta_{1}+\beta_{2}}, 
\beta_{1}<-\frac{\beta_{2}}{4}, \beta_{2}<0, 
\frac{-4 \beta_{2}^{2}-12 \beta_{1}-5 \beta_{2}}{2 \left(2 \beta_{2}^{2}+3 \beta_{1}+4 \beta_{2}\right) \left(4 \beta_{1}+\beta_{2}\right)}<\lambda
\right\}.\label{eq:S123}
\end{align}
None of the previous sets of stability allows for $\beta_1$ and $\beta_2$ to be positive at the same time. In particular, $\beta_2$ is never positive, as it can be seen by combining the first three conditions \eqref{eq:cstr1}-\eqref{eq:cstr3}. 
Examples of solutions in all these three sets of allowed parameter space are presented in Figs.~\ref{fig:exp_&_S1} and~\ref{fig:S2_&_S3}.

\subsection{Stability}

One should note that the solution of $H$ is the full solution, valid without approximations. As for the dynamical equation for the variable $\bar{Y}$, or $A_0^2/\Mpl^2$, we have only solved the solution asymptotically, but we have not linearized it. The equation of motion for $Y$ is already linear, without any approximation. This means that if we perturb $Y$, by $Y+\epsilon\delta Y$, with $|\epsilon\delta Y|\ll |Y|$ as initial condition, then the function $\delta Y$ will satisfy the homogenous solution of Eq.~\eqref{eq:barY_prime}, namely
\begin{align}
\bar{\delta Y}' \left(2\beta_{1}+\beta_{2}\right) \sqrt{-12 \beta_{1}-3 \beta_{2}}\, \tanh \gamma(\tau)=
    \bar{\delta Y}\! \left(4\beta_{1}+\beta_{2}\right)
    -2\left(\beta_{1}+ \beta_{2}\right) \bar{\delta Y} \tanh^{2}\gamma(\tau)\,,
    \label{eq:dyn_fps_stab}
\end{align}
or
\begin{equation}
-\sqrt{-12\beta_{1}-3\beta_{2}}\,\bar{\delta Y}'\,(2\beta_{1}+\beta_{2})\approx\left(\beta_{2}-2\beta_{1}\right)\bar{\delta Y}\,.\label{eq:exp_gro_stab}
\end{equation}
This implies that the homogeneous solutions of Eq.~\eqref{eq:exp_gro} and the ones of Eq.~\eqref{eq:exp_gro_stab} are the same. In particular, the exponential suppression or growth will be reproduced for the same choice of parameters present in the background solutions. This implies that $|\delta Y|/Y$ will remain in general bounded.

\section{The anisotropic universe}

In section {2}, we have seen that the non-minimal coupling induces an additional degree of freedom if the temporal component of the vector field is not vanishing. While this mode is not necessarily a ghost dof, it has a vanishing speed of propagation. This indicates a strong coupling in the theory, meaning that the higher-order terms might be important\footnote{{In addition to the other type of strong coupling, characterized by the vanishing of the kinetic term.}}. One possibility is that this might be related to the isotropy of the space-time. Thus, in this section, we will study the modes in the anisotropic Universe, described by the Bianchi Type I background, which is given by:

\begin{equation}
    ds^2=-N^2(t)dt^2+a^2(t)dx^2+b^2(t)\delta_{ij}dx^idx^j, \qquad i=1,2, \qquad j=1,2
\end{equation}
One should note that it is a term such as $2C(t)dtdx$ is also allowed in the above metric. However, as it can be removed by a coordinate transformation $x=X+F(t)$, with the choice of $F$ such that $C(t)$ is canceled, we will not consider such a term. 

In contrast to the FLRW Universe, where a single vector field could have a non-vanishing value only for the temporal component $A_0(t)$, this case admits a more general solution: 
\begin{equation}
    A_{\mu}=\left(-N(t)A_0(t), A_1(t), 0, 0\right). 
\end{equation}

Similarly to the previous section, we will first find the background equations for the vector field and the metric. Then, we will vary around them, and investigate the degrees of freedom. 

The background equations of motion are found by varying with respect to the lapse, the two scale factors, and the components of the vector field. In particular, by varying with respect to $N(t)$, solving for $\Lambda$, and setting $N=1$, we find: 
\begin{equation}\label{NvarBT1}
 \begin{split}
        \Lambda=&\frac{1}{2 a^{2} \Mpl^{2}}\left[-2 H_{b}^{2} \beta_{1} A_{1}^{2}+4 A_{1} \beta_{1} \left(A_{1} H_{a}-2 \dot{A}_1\right) H_{b}+4 A_{1}^{2} \left(\beta_{1}+\frac{\beta_{2}}{2}\right) H_{a}^{2} \right.\\&\left.-4 \dot{A}_1 A_{1} \left(\beta_{1}+\frac{\beta_{2}}{2}\right) H_{a}+m^{2} A_{0}^2a^2-m^{2} A_{1}^{2}-\dot{A}_1^{2} +\left( \beta_{1}+\frac{\beta_{2}}{2}\right)\left(2H_{b}+H_a\right)4\dot{A}_0 A_{0}
       \right.\\ &\left. + a^{2}\left(2 A_{0}^{2} \beta_{1}+2 A_{0}^{2} \beta_{2}+2 \Mpl^{2}\right) \left(H_{b}^{2}+2H_aH_b\right) \right]
 \end{split}
\end{equation}
Here, $H_a=\frac{\dot{a}}{a}$ and $H_b=\frac{\dot{b}}{b}$ stand for the Hubble constants for the scale factors $a(t)$ and $b(t)$, while the dot denotes the derivative with respect to time. From now on, we will present the results with $N=1$. By varying with respect to $A_0$, we find: 
\begin{equation}
    4 b A_{0} \left[\left(\beta_{1}+\frac{\beta_{2}}{2}\right)\left(\frac{b\Ddot{a}}{2}+a\Ddot{b}\right)+\frac{ab}{2} \left(2 H_{a} H_{b}+ H_{b}^{2}\right) \beta_{1}+\frac{m^2a}{4} \right]=0.
\end{equation}
Similarly to the previous section, there are two possibilities for $A_0(t)$ -- it can have a vanishing and a non-vanishing value. Since we are interested in the rise of the extra dof, we will consider only the case when the temporal component has a non-vanishing value.\footnote{However, later on, we will look into the limit of small values of $A_0/\Mpl$ and the behavior of the perturbations in this regime.} The variation with respect to $A_1(t)$ yields: 
\begin{equation}
    \begin{split}
      \Ddot{A}_1 =& -\frac{ A_{1} \left(2 \beta_{1}+\beta_{2}\right) \Ddot{a}}{2a}-4 \frac{\Ddot{b}}{b} A_{1} \beta_{1}-2 \left(\left(2 H_{a} H_{b}+H_{b}^{2}\right) \beta_{1}+H_{a} H_{b} \beta_{2}+\frac{m^{2}}{2}\right) A_{1}\\&-\dot{A}_1 \left(H_{a}-2 H_{b}\right) 
    \end{split}
\end{equation}
By varying the action with respect to the first scale factor $a(t)$ we find: 
\begin{equation}\label{avarBT1}
    \begin{split}
   &2 \left(\left(A_{0}^{2} \beta_{1}+A_{0}^{2} \beta_{2}+\Mpl^{2}\right) a^{2}+A_{1}^{2} \beta_{1}\right) a \Ddot{b}+b \left(\beta_{1}+\frac{\beta_{2}}{2}\right)\left(2 A_{0}a^{3} \Ddot{A}_0-2 a A_{1} \Ddot{A}_1+4A_{1}^{2} \Ddot{a}\right)\\& +ba^{3}\left(A_{0}^{2} H_{b}^{2}+4 A_{0} \dot{A}_0 H_{b}+2 \dot{A}_0^{2}\right) \beta_{1}+\left(A_{0}^{2} H_{b}^{2}+4 A_{0} \dot{A}_0 H_{b}+\dot{A}_0^{2}\right) ba^{3}\beta_{2}-\Lambda \,\Mpl^{2}a^{3}b\\&+\frac{m^{2}a^{3} A_{0}^{2}b}{2}+a^{3}b\Mpl^{2} H_{b}^{2}+\left(\left(-6 H_{a}^{2}+8 H_{a} H_{b}+H_{b}^{2}\right) \beta_{1}+\left(-3 H_{a}^{2}+4 H_{a} H_{b}\right) \beta_{2}\right) abA_{1}^{2}\\&+\frac{m^{2}}{2}abA_{1}^{2}-4ba \dot{A}_1 \left(H_{b}-2 H_{a}\right) \left(\beta_{1}+\frac{\beta_{2}}{2}\right) A_{1}-2 ab\dot{A}_1^{2} \left(\beta_{1}+\frac{\beta_{2}}{2}+\frac{1}{4}\right) =0
    \end{split}
\end{equation}

By solving the equations (\ref{NvarBT1})-(\ref{avarBT1}) for $\Lambda, \Ddot{a}, \Ddot{A}_0$ and $\Ddot{A_1}(t)$, and substituting them to the equation obtained by varying the action with respect to $b(t)$, we find the last equation: 
\begin{equation}
    \begin{split}
      \Ddot{b}=& -\frac{b}{3 \left(\beta_{1}+\frac{\beta_{2}}{2}\right) \left(-\frac{2 A_{1}^{2} \beta_{2}^{2}}{3}+\left(a^{2} A_{0}^{2}-\frac{4 A_{1}^{2}}{3}\right) \beta_{2}+\left(a^{2} A_{0}^{2}-A_{1}^{2}\right) \beta_{1}+\Mpl^{2} a^{2}\right)}\times\\&\left[\beta_{2}^{3}H_{a}  H_{b}A_{1}^{2} +2\beta_{2}^{2}\beta_{1} H_{a}H_{b} A_{1}^{2}  -\frac{\beta_{2}^{2} a^{2}}{2}A_{0} \left(H_{a}-H_{b}\right) \left(A_{0} H_{b}+2 \dot{A}_0\right)+\dot{A}_1 A_{1} H_{a}\right.\\ &\left.-A_{1}^{2} H_{a}\beta_{2}^{2} \left(H_{a}-2 H_{b}\right)-\frac{\dot{A}_1^{2}}{2}\beta_{2}^{2}+\frac{\beta_{2}\beta_{1}}{2}\left(5 A_{0} H_{b}^{2}+\left(A_{0} H_{a}+6 \dot{A}_0\right) H_{b}-6 \dot{A}_0 H_{a}\right) A_{0} a^{2}\right.\\ &\left.+\beta_{2}\beta_{1}\left(-3 H_{a}^{2}-\frac{3}{2} H_{b}^{2}+\frac{7}{2} H_{a} H_{b}\right) A_{1}^{2}+3 \beta_{2}\beta_{1}\left(H_{a}-\frac{H_{b}}{3}\right) \dot{A}_1 A_{1}-\beta_{2}\beta_{1}\dot{A}_1^{2}-\beta_{1}\dot{A}_1^{2}\right.\\ &\left.+\frac{\beta_{2}}{2}\left(-\Mpl^{2} H_{a} H_{b}+\Mpl^{2} H_{b}^{2}+m^{2} A_{0}^{2}\right) a^{2}-\beta_{2}\frac{\dot{A}_1^{2}}{2}-\frac{m^{2} A_{1}^{2}}{2}\beta_{2}+\frac{\Mpl^{2} a^{2} m^{2}}{2}-\frac{\beta_{1}m^{2} A_{1}^{2}}{2}\right.\\ &\left.+\beta_{1}^{2}\left(2 A_{0} H_{b}^{2}+\left(A_{0} H_{a}+2 \dot{A}_0\right) H_{b}-2 \dot{A}_0 H_{a}\right) A_{0} a^{2}+2 \beta_{1}^{2}A_1\left(H_{a}-H_{b}\right) \dot{A}_1\right.\\ &\left.+2 \beta_{1}^{2}\left(-H_{a}^{2}+\frac{1}{2} H_{a} H_{b}-H_{b}^{2}\right) A_{1}A_{1}+ \beta_{1}\left(2 \Mpl^{2} H_{b}^{2}+\frac{1}{2} m^{2} A_{0}^{2}+\Mpl^{2} H_{a} H_{b}\right) a^{2}\right]
    \end{split}
\end{equation}

This equation can be further substituted into the previous ones for $\Ddot{b}$, resulting in the expressions for the scale factors $\Ddot{a}$ and $ \Ddot{b},$ the cosmological constant$, \Lambda,$ and the background values of the vector field $ \Ddot{A}_0$ and $\Ddot{A}_1$.  In contrast to the FLRW case, the background equations are much more complicated, and further analysis is necessary to obtain the general conditions under which there is an accelerated expansion along the x or $\{y, z\}$ axis. For our purposes, it will be sufficient to assume that they are satisfied when studying the action.   

As a next step, let us perturb the metric and the vector field around the background values: 
\begin{equation}
    g_{\mu\nu}= g_{\mu\nu}^{(0)}+\delta g_{\mu\nu}\qquad \qquad A_{\mu}=A_{\mu}^{(0)}+\delta A_{\mu}
\end{equation}
The vector field perturbations can be written as: 
\begin{equation}
    \begin{split}
        & \delta A_0=-A_0 A\\
        & \delta A_1=A_1 \pi_{,x}\\
        & \delta A_i=-A_i^T+\chi_{,i}\qquad A_i^T=\varepsilon_{ij}v_{A,j}.
    \end{split}
\end{equation}
The metric perturbations are given by: 
\begin{equation}
    \begin{split}
        &\delta g_{00}=2\phi\\
        &\delta g_{0x}=\omega_{,x}\\
        &\delta g_{oi}=v_i+B_{,i},\qquad v_i=\varepsilon_{ij}v_{,j}\\
        &\delta g_{xx}=a^2\psi\\
        &\delta g_{xi}=\lambda_{i,x}+\mu_{,xi}\qquad  \lambda_i=\varepsilon_{ij}\lambda_{,j}\\
        &\delta g_{ij}=b^2(2\tau \delta_{ij}+2E_{,ij}+h_{i,j}+h_{j,i}), \qquad h_i=\varepsilon_{ij}h_{,j}
    \end{split}
\end{equation}
In the above relations, we have set $N=1$. Furthermore, $\varepsilon_{ij}$ is the Levi-Civita symbol, with $\varepsilon_{yz}=1$, and $,i$ denotes a derivative with respect to $\{y, z\}$. The modes $\phi, \omega, B, \psi, \mu, \tau, E, A, \pi, \chi$ are called even modes, while $v, \lambda, h, v_A$ are called the odd modes. In linearized order the even and odd modes decouple, allowing us to study them separately. 

\subsection{Gauge-invariant variables }
Under the infinitesimal coordinate transformations, 
\begin{equation}\label{infcoo}
    x^{\mu}\to\Tilde{x}^{\mu}=x^{\mu}+\xi^{\mu}, 
\end{equation}
with 
\begin{equation}
    \begin{split}
        \xi^0\qquad\qquad
        \xi^1_{,x}\qquad\qquad
        \xi_i=\delta_{ij}\xi^j=\zeta_{,i}+\varepsilon_{ij}\xi_{,j}
    \end{split}
\end{equation}
we find that the even modes transform as: 
\begin{equation}
    \begin{split}
        &\tilde{\phi}=\phi+\dot{\xi}^0\\
    &\tilde{\omega}=\omega+\xi^0-a^2\dot{\xi}^1\\
    &\tilde{B}=B+\xi^0-b^2\dot{\zeta}\\
    &\tilde{\psi}=\psi-2\frac{\dot{a}}{a}\xi^0-2\xi^1_{,xx}\\
    &\tilde{\mu}=\mu-a^2\xi^1-b^2\zeta\\
    &\tilde{\tau}=\tau-\frac{\dot{b}}{b}\xi^0\\
    &\tilde{E}=E-\zeta\\
    &\tilde{A}=A-\frac{1}{A_0}\left[(A_0\xi^0)^{.}-A_1\dot{\xi}^1_{,x}\right]\\
    &\tilde{\pi}_{,x}=\pi_{,x}+\frac{1}{A_1}\left[A_0\xi^0_{,x}-\dot{A}_1\xi^0-A_1\xi^1_{,xx}\right]\\
    &\tilde{\chi}=\chi+A_0\xi^0-A_1\xi^1_{,x}
    \end{split}
\end{equation}
With the help of the above transformation laws, we can define the following gauge-invariant variables for the even modes: 
\begin{equation}
    \begin{split}
        &\Phi=\phi+\left(\frac{b\tau}{\dot{b}}\right)^{.}\\
        &\Omega=\omega+\frac{b\tau}{\dot{b}}\tau-a^2\left[\frac{1}{a^2}\left(\mu-b^2E\right)\right]^{.}\\
        &\underline{B}=B+\frac{b\tau}{\dot{b}}\tau+b^2\dot{E}\\
        &\Psi=\psi-2\frac{\dot{a}b}{a\dot{b}}\tau-2\partial_x^2(\mu-2b^2E)\\
        &\underline{A}=A-\frac{1}{A_0}\left[\left(A_0\frac{b\tau}{\dot{b}}\right)^{.}-A_1\partial_x\left(\frac{1}{a^2}(\mu-b^2 E)\right)^{.}\right]\\
        &\underline{\chi}=\chi+A_0\frac{b}{\dot{b}}\tau-A_1\partial_x(\mu-b^2 E)\\
        &\partial_x\underline{\pi}=\partial_x\pi+\frac{1}{A_1}\left[A_0\partial_x\left(\frac{b}{\dot{b}}\tau\right)-\dot{A}_1\frac{b}{\dot{b}}\tau-A_1\partial_x^2(\mu-b^2E)\right]
    \end{split}
\end{equation}
The odd modes transform under the infinitesimal coordinate transformation as:
\begin{equation}
    \begin{split}
        &\tilde{h}=h-\xi\\
        &\tilde{v}=v-b^2\dot{\xi}\\
        &\tilde{\lambda}=\lambda-b^2\xi\\
        &\tilde{v}_A=v_A
    \end{split}
\end{equation}
We can notice that $v_A$ is already gauge invariant. The remaining two gauge invariant variables can be then defined as: 
\begin{equation}
    \begin{split}
        V=v-b^2\dot{h}\qquad\text{and}\qquad
        \underline{\lambda}=\lambda-b^2h
    \end{split}
\end{equation}
In the following, we will analyze the perturbations around the Bianchi Type I background. One can easily check that the odd and even modes decouple for the linearized theory, allowing us to study them separately, similarly to the scalar, vector, and tensor modes in the homogeneous and isotropic background.  Due to the complicated expressions, we will first describe the procedure to analyze the modes and write the theory only in terms of the propagating ones. Then, we will study the behavior of these modes for several choices of parameters, to study the no-ghost conditions and their speed of propagation. 

\subsection{Odd modes }
Let us first consider the odd modes, which yield much easier expressions than the even ones. We will work in the gauge: 
\begin{equation}
    h=0. 
\end{equation}
In this case, the modes $v_A, v $ and $\lambda$ coincide with the gauge-invariant variables we have found in the previous subsection. By expanding the action up to the second order in perturbation and performing several integrations by parts, we find the following expression for the Lagrangian density: 

\begin{equation}
   \begin{split}
        \mathcal{L}&= k_1\dot{v_A}\Delta \dot{v_A}+k_2\Delta v \lambda''+ k_3\Delta v\Delta v+ k_4\Delta v_A\Delta v_A+ k_5\Delta v v+k_6\Delta v_A v+k_7\Delta v_A v_A\\&+k_8\Delta\dot{v}v+k_9\Delta v'v+k_{10}\Delta v'v_A+k_{11}\Delta v'v'+k_{12}\Delta v_A'v'+k_{13}\Delta \lambda'v'+k_{13}\Delta \lambda'v_A
   \end{split}
\end{equation}
Here, the Laplacian stands for $\Delta=\partial_i\partial_i$ with $i=y,z$, the prime denotes the derivative with respect to x coordinate, while the dot denotes the derivative with respect to the coordinate time $t$. The time-dependent coefficients $k_1 ... k_{14}$ are given in the Appendix \ref{APPodd}. 
As a next step, we will study the modes in the Fourier space:
\begin{equation}
    X=\int \frac{dkd^2q}{(2\pi)^{3/2}}X(t,k,q)e^{ikx+iq_iy^i}, \qquad y^i={y,z},
\end{equation}
where $X$ stands for the odd modes. By studying the above Lagrangian density, we can notice that the mode $v$ does not show up with any time derivative. As a corresponding check, one can verify that the determinant of the kinetic matrix made out of the modes is vanishing. Therefore, similarly to the analysis in the FLRW background, we will find the constraint for the mode $v$. By solving it, and substituting it back to the action, we find a Lagrangian density that only depends on the two remaining modes, $\lambda$ and $v_A$. However, both of them now have non-trivial kinetic terms, and the determinant of the corresponding kinetic matrix does not vanish in this case. Therefore, in the odd sector, we find two propagating degrees of freedom. 
In the following, we will first analyze the even modes, and then discuss the no-ghost conditions and the speed of propagation for all of the modes.

\subsection{Even modes }
Let us {now} study the even modes. We will work in the following gauge: 

\begin{equation}
    E=0,\qquad \tau=0,\qquad\text{and}\qquad \mu=0. 
\end{equation}
By expanding the action (\ref{NMProcaAction}) to the second order in perturbations, and performing several integrations by parts, we find the following Lagrangian density {for the even modes}:

\begin{equation}\label{bianchiLProca}
    \begin{split}
\mathcal{L}&=f_1\dot{\psi}^2+f_2\dot{\psi}\dot{\phi}+f_3\dot{\psi}\dot{A}+f_4\dot{\psi}\dot{\pi}'+f_5\dot{\omega}'\dot{\psi}+f_6\dot{\pi}'' \dot{\pi}+f_7\Delta\dot{\chi} \dot{\chi} \\& 
        +f_8\Delta\dot{\psi} B+f_9\Delta\dot{\psi} \chi+f_{10}\Delta\dot{\phi} B+f_{11}\dot{\pi}'' A+f_{12}\Delta\dot{\chi} A+
f_{13}\dot{\pi}'' \pi+ f_{14}\dot{\omega}'' A +f_{15}\dot{\phi}'' \omega\\&+
f_{16}\dot{\omega}'' \omega+f_{17}\dot{\omega}'' \phi+f_{18}\dot{\omega}'' \psi+
+f_{19}\Delta\dot{B} B+f_{20}\Delta\dot{B} A+f_{21}\Delta\dot{B} \phi+f_{22}\Delta\dot{B} \psi+
+f_{23}\dot\phi A\\&+f_{24}\dot\psi A+f_{25}\dot{\phi} \psi+f_{26}\dot{\psi} \phi+ f_{27}\dot{\phi}'\pi+f_{28}\dot{\phi}'\omega +f_{29}\dot{\psi}'\omega +f_{30}\dot{\psi}'\pi+f_{31}\phi\dot{\pi}'+f_{32}\psi\dot{\pi}'
\\&+ f_{33}\phi'\pi +f_{34}\phi' A+ \phi' \omega+ f_{35}\psi' A+ f_{36}\psi' \pi + f_{37}\psi'\omega +f_{38}\psi'\phi+ f_{39}\chi'\Delta\chi'+f_{40}\omega' A
\\&+ f_{41}AA''+ f_{42}A''\pi+f_{43}\pi''\pi+ f_{44}\omega''A+f_{45}\omega''\pi + f_{46}\omega''\omega
          \\&+ f_{47}\phi'' A+f_{48}\phi'' \pi+f_{49}\phi'' \omega+f_{50}\phi'' \phi+f_{51}\psi''\omega+f_{52}\psi''\phi+ 
          \\& +f_{53}\chi'_{,i}\pi'_{,i}+f_{54}\omega'_{,i}B'_{,i}+f_{55}B\Delta B+ f_{56}A\Delta B+f_{57}\chi\Delta B+f_{58}A\Delta A+ f_{59}\chi \Delta\chi 
         \\& +f_{60}\Delta B' B +f_{61}\Delta B' A+f_{62}\Delta B' \pi+f_{63}\Delta B' \chi+ f_{64}\Delta B' \omega+ f_{65}\Delta B'  B'\\&+f_{66}\Delta \pi' B+
f_{67}\Delta \pi' \pi'+f_{68}\Delta\chi'B
+ f_{69}\omega''' \pi+ f_{70}\omega''' \omega+f_{71}\phi''' \pi+ f_{72}\phi''' \omega +  f_{73}\Delta\omega' A\\&+f_{74}\Delta\omega' \chi+f_{75}\Delta\omega' \omega'
+ f_{76}\Delta \phi' \pi+f_{77}\Delta \phi' \chi+ f_{78}\Delta \phi' \omega+f_{79}\Delta \psi' \pi+ f_{80}\Delta \psi' \omega+
       \\&+ f_{81}\Delta\phi B+f_{82}\Delta\phi A+f_{83}\Delta\phi \chi+f_{84}\Delta\phi \phi +f_{85}\Delta\psi B+f_{86}\Delta\psi A\\&+f_{87}\Delta\psi \chi+f_{88}\Delta\psi \phi+f_{89}\Delta\psi \psi+f_{90}A^{2}+ f_{91}\phi^{2}+f_{92}\psi^{2}+f_{93}\phi A + f_{94}\phi\psi + f_{94}\psi A+
    \end{split}
\end{equation}
Here, the functions $f_1 ... f_{94}$ are time-dependent functions of scale factors and background values for the vector field. They are given explicitly in the Appendix \ref{APP2}. 

{Similarly to the previous section, in order to study the propagating dof, we will express the fields in terms of the Fourier modes: }
\begin{equation}
    X=\int \frac{dkd^2q}{(2\pi)^{3/2}}X(t,k,q)e^{ikx+iq_iy^i}, \qquad y^i={y,z}
\end{equation}
{Here, $X$ now stands for all of the even modes. In this subsection, we will describe the procedure with which one can find the propagating modes of the theory. Their stability will be then studied in the following subsections. }

By studying the action (\ref{bianchiLProca}), and assuming that the background equations of motion are satisfied, we can notice that among seven fields, $\phi, \omega, B$ and $A$ do not come with time derivatives. Since our goal is to express the action only in terms of the propagating modes, first we find the constraint for the field $B$, solve it, and substitute it back to the action. The resulting action is then only a function of six fields.  One can find that all fields appear with time derivatives at this point. However, the determinant of the kinetic matrix vanishes, thus indicating that not all fields are propagating.  By substituting,
\begin{equation}
    \omega=\omega_2+\frac{ia^2A_0}{kA_1}\left(\phi+A\right)
\end{equation}
we can find that the fields $\phi$ and $A$ are not propagating. We then find a constraint for $A$, solve it, and substitute it back into the action. By further inspecting the action, one can notice that $\phi$ is not propagating. By finding its constraint, solving it and substituting it back to the action, one obtains an expression that contains only four fields, all of which are now propagating -- $\psi, \omega_2, \chi$, and $\pi$. 
Since the determinant of the corresponding kinetic matrix is non-vanishing, this leads us to the conclusion that all four fields are propagating dof.

{Therefore, we have overall found four propagating even modes, and two odd modes. The number of the dof matches also the homogeneous and isotropic case, now with much more complicated expressions. Due to their complexity, we will have to study them only for particular values of the parameters of the theory. This will be done in the following section. }

\subsection{Results of stability conditions of Bianchi-I }

In the above,
we were able to integrate out all the auxiliary fields, leaving a reduced Lagrangian with four perturbation fields for which its kinetic matrix has a non-zero determinant, in general, {for the even modes}. {In addition, we have found a reduced Lagrangian with two perturbation fields in the odd sector.} More precisely, the {total} reduced Lagrangian density {containing both even and odd modes} can be cast into the following form
\begin{align}
\mathcal{L} & =A_{ij}(t,k,q)\dot{\psi}_{i}(t,-k,-q)\,\dot{\psi}_{j}(t,k,q)-M_{ij}(t,k,q)\,\psi_{i}(t,-k,-q)\,\psi_{j}(t,k,q)\nonumber\\
&+B_{ij}(t,k,q)\,[\dot{\psi}_{i}(t,k,q)\psi_{j}(t,-k,-q)-\dot{\psi}_{j}(t,-k,-q)\psi_{i}(t,k,q)]\,,
\end{align}
where the indices $i$ and $j$ run from one to {six}%four 
and by $\psi_i$, we mean the fields that remained after integrating out all the auxiliary ones. More specifically, $\vec{\psi}=(\psi, \omega_2, \chi, \pi, \lambda, v_A)$. 
The matrix elements $A_{ij}$, $B_{ij}$ and $M_{ij}$ satisfy the
following relations
$
A_{ji}^{*}  =A_{ij}\,,
B_{ji}^{*}  =-B_{ij}\,,
M_{ji}^{*}  =M_{ij}\,,$
making the matrices $\bf{A}$ and $\bf{M}$ hermitian and the matrix $\bf {B}$ anti-hermitian.\footnotetext{In terms of $k$ and $q$, we find
$A_{ji}(t,-k,-q)  =A_{ij}(t,k,q)$, $B_{ji}(t,-k,-q)  =-B_{ij}(t,k,q)$, and $M_{ji}(t,-k,-q)  =M_{ij}(t,k,q)$.}

Since we have integrated out all the auxiliary modes, one finds that $\det{\bf{A}}\neq0$, in general. For this system of multi-field, the equations of motion, in WKB approximation, become the following ones
\begin{equation}
-A_{ij}(t,k,q)\ddot{\psi}_{j}(t,k,q)+2\dot{\psi}_{j}(t,k,q)B_{ji}(t,k,q)-M_{ij}(t,k,q)\psi_{j}(t,k,q)\approx0\,,
\end{equation}
for which we need to find the discriminant equation
\begin{equation}
\det[\omega^{2}A-2i\omega B^{T}-M]=0\,.
\end{equation}
This eighth-order algebraic equation has to be solved in terms of $\omega$. Once, we have these eight solutions, we can distinguish the group velocity of the modes in terms of their direction of propagation. The modes may propagate along the $x$-direction, on which we can define $c_x=a\partial\omega/\partial k$, or propagate in the space-isotropic $y$-$z$ plane, according to $c_{\rm yz}=b\partial\omega/\partial q$. Evidently for these speeds to be sensible, one requires that they do not become complex, otherwise, in general, a fast-growing instability would take place.

The study of the discriminant equation turns out to be complicated as the matrix elements are complicated functions. We found it convenient to study them in a way that allowed us to shed some light on the behavior of such an equation. First of all, by using the background equations, it is possible to remove from each matrix coefficient the second and higher time derivatives of the background quantities. At this point each matrix coefficient is going to be a function of $a, b, H_a, H_b, A_0, A_1, \dot{A}_0, \dot{A}_1$, of $k$ and $q$, and the parameters $\beta_1,\beta_2,m,\Mpl$.\footnote{Here, we assume having used the equations of motion to substitute $\ddot{a}$, $\ddot{b}$ (and possibly their higher time derivatives), and $\Lambda$ in terms of all the other variables.} We have then chosen to set the parameters so that the solution belongs to one of the FLRW-compatible intervals of stability, namely $\mathcal{S}_1$, $\mathcal{S}_2$, and $\mathcal{S}_3$. Next, we focused our attention on some choices of $a, b, H_a, H_b, A_0, A_1, \dot{A}_0, \dot{A}_1$ which were close to the isotropic stable fixed points of FLRW. In particular by setting numerical values for each of the previous background quantities compatible with the following quasi-isotropic constraints: $a\approx b$, $H_a\approx H_b\approx H_{\rm dS}$, $A_0/\Mpl\approx \sqrt{\bar{Y}_{\rm fin}} $, $|A_1|/\Mpl\ll1$, and $|\dot{A_1}|/(m\Mpl)\ll1$. We will perform this analysis in the upcoming material.

\subsection{Discussion of some different cases }

Here we describe the behavior of the discriminant equation under some assumptions. Let us then discuss the following cases.
\begin{enumerate}
    \item Example belonging to the set $\mathcal{S}_1$. For instance, we can take $\beta_1=-1$, $\beta_2=-4$, {$\Lambda=1/9$}%$\lambda=1/9$
    , $A_0=10$, $\Mpl=30$, $m=1$, $a=1+10^{-2}$, $b=1-10^{-2}$, $\dot{a}=1/\sqrt{24}(1+10^{-2})$, $\dot{b}=1/\sqrt{24}(1-10^{-2})$, $A_1=10^{-2}$, $\dot{A}_1=10^{-2}$, and $\dot{A}_0=10^{-2}$.
    \item Example belonging to the set $\mathcal{S}_2$. For instance, we can take $\beta_1=-1/30$, $\beta_2=-1/10$, {$\Lambda=4$}%$\lambda=4$
    , $A_0=10$, $\Mpl=5$, $m=1$, $a=1+10^{-2}$, $b=1-10^{-2}$, $\dot{a}=\sqrt{10/7}(1+10^{-3})$, $\dot{b}=\sqrt{10/7}(1-10^{-3})$, $A_1=10^{-3}$, $\dot{A}_1=10^{-3}$, and $\dot{A}_0=10^{-3}$.
    \item Example belonging to the set $\mathcal{S}_3$. For instance, we can take $\beta_1=1/30$, $\beta_2=-3/10$, {$\Lambda=3$}%$\lambda=3$
    , $A_0=10$, $\Mpl=10\sqrt{11/30}$, $m=1$, $a=1+10^{-4}$, $b=1-10^{-4}$, $\dot{a}=\sqrt{2}(1+10^{-4})$, $\dot{b}=\sqrt{2}(1-10^{-4})$, $A_1=10^{-4}$, $\dot{A}_1=10^{-4}$, and $\dot{A}_0=10^{-4}$.
\end{enumerate}

In the previous three examples, although all close to the isotropic fixed points, there are some wanted deviations from pure isotropy. We have considered different levels of isotropy deviation from a residual of $10^{-4}$ up to $10^{-2}$. In all the previous cases, we find results having the same following pattern:

{\textbf{{Even modes:} }}
\begin{itemize}
    \item In all cases, one no-ghost condition is proportional to $( \mathcal{A} k^2- \mathcal{B} q^2)^2$, with here and in the following $\mathcal{A}$ and $\mathcal{B}$ being positive numerical constants, whose value is not important to report here. This indicates a strong coupling for some modes with a determinate value of $k^2/q^2$. 
    \item In the $x$-direction, one of the modes propagates with $\omega^2 = \mathcal{A}\,k^2/a^2$ ($\mathcal{A}>0$), suggesting a modified dispersion relation. Solutions instead with $\omega=c_x k/a$, have the following structure: two modes (four solutions for $\omega$) are stable, whereas the last mode gives this $(c_s-\mathcal{A})^2$, which indicates a propagation in the only one $x$-direction.
    \item In the isotropic space, {that is studying the propagation for the case $q/b\gg k/a\gg \max(H_a, H_b)$}, the discriminant equation gives a mode propagating with 0 speed of propagation, indicating a second strong coupling. Another mode is stable and the two remaining modes are unstable. {It is important to note that the speed of propagation in this case is $k-$dependent with $k$ now being the parameter. Therefore, to recover the above results, one should consider particular large values for $k$. }
\end{itemize}

{\textbf{{Odd modes:} }}
\begin{itemize}
    \item In all cases, the no ghost conditions are satisfied. 
    \item In the $x$-direction, the speeds of propagation for the modes are non-vanishing. Interestingly, for each mode, they appear in the form $(c_{odd,1}-A)(c_{odd,2}+B)$ where the coefficients $A$ and $B$ slightly differ, instead of the standard form $(c_{odd, 1,2}^2-A)$, leading to four different speeds of propagation, two associated with each mode.\footnote{{For example, the difference in the speeds of propagation in the x direction for the first example is $1.730072731, -1.734033129, 1.413224070, -1.415204268$.}}
    \item In contrast to the $x$-direction, the isotropic case for all examples contains speeds of propagation in the standard form: $(c_{odd}^2-A)$  for each of the two modes. 
  
\end{itemize}

Although the above structure holds for those examples, we point out that we have not scanned the whole parameter space for the parameters.\footnote{{For fixed point solutions in the case $m^2<0$, we have verified that one mode propagates in only one direction, for another one $c_s^2=0$ vanishes in the isotropic subspace, where two modes are unstable for {the even perturbations}.}} Nonetheless, we have also looked into the following particular cases, {for the even modes}. 

\begin{enumerate}
    \setcounter{enumi}{3}
    \item Case with $2\beta_1\approx\beta_2$, as to try to have large values of $A_0/\Mpl$. An example for this case belonging to the set $\mathcal{S}_1$, is the following: $\beta_1=-2+10^{-4}$, $\beta_2=-4$, and in this case, $\Lambda$ is constrained to belong to a tiny interval, leading to effective fine-tuning.\footnote{In fact, $5000/59999<\Lambda<2500/29999$.} Choosing $\Lambda={\frac{149996250}{1799910001}}$, $\Mpl=1$, $m=1$, $a=1+10^{^4}$, $b=1-10^{^4}$, $\dot{A}_0=10^{-4}=\dot{A}_1$, $A_1=10^{-4}$, $\dot{a}=\sqrt{2500/89997}\,(1+10^{-4})$,$\dot{b}=\sqrt{2500/89997}\,(1-10^{-4})$, from which we determine $A_0=50\sqrt{119998}/59999\approx0.288$, which is not that large after all, because of the $\Lambda$-tuning. The results for this case are not different from the ones found for cases 1, 2, or 3.

    \item Small $A_0/\Mpl$ cases. We have looked into this case by setting $\beta_1=-1$, $\beta_2=-4$, $\Lambda=1/8-10^{-8}$, $\Mpl=1$, $m=1$, $a=1+10^{^4}$, $b=1-10^{^4}$, $\dot{A}_0=10^{-4}=\dot{A}_1$, $A_1=10^{-4}$, $\dot{a}=1/\sqrt{24}\,(1+10^{-4})$,$\dot{b}=1/\sqrt{24}\,(1-10^{-4})$, and $A_0=\sqrt{2}/5000$. As for the no-ghost condition, the $k/q$-dependent vanishing no-ghost condition is still present. In this same case, in the $x$-propagating direction, beside the mode with modified dispersion relation and the two equal $c_s$-solutions, (i.e.\ satisfying $(c_s-\mathcal{A})^2$) mode, one mode is stable whereas the last one becomes unstable. In the $y{-}z$ plane propagation, besides the mode with $c_s^2=0$, we find the other three modes unstable. 
    \item Looking for a large positive value of the cosmological constant. To find one example for this case, one can pick up in the $\mathcal{S}_1$ set the following combination: $\beta_1=1-10^{-4}$, $\beta_2=-4$, $\Lambda=2499$, $A_0=2/\sqrt{29999}$, $\Mpl=1$, $m=1$, $a=1+10^{^4}$, $b=1-10^{^4}$, $\dot{A}_0=10^{-4}=\dot{A}_1$, $A_1=10^{-4}$, $\dot{a}=\sqrt{2500/3}\,(1+10^{-4})$,$\dot{b}=\sqrt{2500/3}\,(1-10^{-4})$. This case effectively leads to a small value for $A_0/\Mpl$, so the results are the same as in the previous case.

\end{enumerate}

{For the odd modes, the above three cases do not significantly change when compared to the first three examples. In particular, all the no-ghost conditions will still be satisfied, and the results for the speed of propagation in the isotropic case still hold. There are some slight changes of the speed of propagation in the x-direction. Namely, in case 5, with small $A_0/\Mpl$, we find that its speed of propagation matches with the isotropic case, and takes the standard form. On the other hand, in case 6, with a large positive value of the cosmological constant, we find that in the x-direction, the speed of propagation for one mode takes the standard form, while for the other takes the   $(c_{odd,1}-A)(c_{odd,2}+B)$ form. }

\section{Conclusion and Summary}

Vector tensor theories are of great interest for several reasons. First of all, a realization of it exists in nature, in terms of the electromagnetic field, which can be described in terms of a four-vector $A_\mu$. The Lagrangian of this theory is $U(1)$ gauge invariant bringing into existence two propagating transverse degrees of freedom, the particle we name by photon. From the pioneering work of Proca, theoretical work was performed to search for vector tensor theories seen as extensions of the standard electromagnetism. Several possibilities were present: 1) to enlarge the theory keeping the same gauge $U(1)$ invariance; 2) to explore different invariance under different gauge groups; 3) abandon the gauge invariance to allow, for instance, the presence of a mass for the photon. When dark energy theories became worthy of investigation, several new theories were introduced in different contexts—for instance, generalized scalar-tensor theories, modifications of gravity, and generalized vector-tensor theories. In particular, the latter ones were built so that only three modes were propagating in the theories, two transverse and one longitudinal mode, besides gravitational waves and the fields coming from the matter sector. The theories were built so that, by construction, the time component $A_0$ of the vector field would not acquire any kinetic term. Being a purely Lagrange multiplier, it could have integrated out from the Lagrangian, leaving the three remaining components of $A_\mu$, to play a non-trivial role in terms of the propagation of the perturbations.

Theories without this ad-hoc construction were also introduced, and these are the ones we have studied in this work. In particular, we studied actions/theories containing in the scalar Lagrangian, besides a standard kinetic term proportional to $F_{\mu\nu}F^{\mu\nu}$, couplings with both the Ricci scalar and the Ricci tensor, in a form proportional to $(m^2+\beta_1  R) A_\mu A^\mu+\beta_2 A_\mu A_\mu R^{\mu\nu}$. For values $\beta_1\neq-\tfrac{1}{2}\beta_2$, the theory does not belong to the generalized Proca action, so we should expect four degrees of freedom to propagate, and indeed this is the case, in general.

We have then first tried to understand better the theory by studying its behavior on an FLRW background in the presence of an Einstein-Hilbert term, a cosmological constant but in the absence of any other matter degree of freedom. Because of the absence of the $U(1)$ gauge symmetry, but just imposing the symmetries of the background, we can have a non-trivial configuration for the vector field as in $A_\mu \,dx^\mu=A_0(t)\,dt$. The theory presents itself as having in general two branches on this background. These branches are generated from the equation of motion of the vector field itself, leading to $[3 \left(\beta_{2}+2 \beta_{1}\right) \dot{H}+3 \left(4 \beta_{1}+\beta_{2}\right)H^{2}+m^{2}] A_{0}=0.$ This solution admits the solution $A_0=0$, and if $A_0\neq0$, it gives a relation between $H$, its derivative, and the parameters of the vector Lagrangian, $m,\beta_1$, and $\beta_2$. The Friedmann equation instead states the dynamics for the field $A_0(t)$. At this point, we studied the background dynamics of the branch $A_0\neq0$. Furthermore, we have studied the perturbation field dynamics for this branch and found the following phenomenology. The background solutions, when constrained by the presence of stable fixed points and stable perturbation dynamics (including the positivity of the no-ghost condition and the squared speed of propagation) then form three different sets of parameter space, that we call $\mathcal{S_1}$, $\mathcal{S_2}$, and $\mathcal{S_3}$. Furthermore, we do find six propagating modes: two gravitational waves, two vectors, and two scalars {in agreement with previous expectations}. 
{Interestingly, we show that the Ricci scalar keeps the speed of propagation for the vector and tensor modes to be unity, while the Ricci tensor deforms this result.} However, {more surprisingly, we find that} the extra scalar mode has two properties: 1) its speed of propagation is zero and 2) its kinetic term can be positive but tends to vanish in the limit {in three different cases: for} $A_0\to0$, {$\beta_1\to-\frac{\beta_2}{2}$ or when both coupling constants tend to zero -- $\beta_1\to0 $ and $\beta_2\to 0$}.

 The first of these two properties seems to indicate that the non-linear terms dominate regardless of the behavior of the homogeneous and isotropic background. If this strong coupling takes place, it indicates that the linearized theory is not trustable in this particular background,  and one needs to find a way to fix or understand this problem. 
To try to understand the behavior of this mode and even the 

{limits, when the kinetic terms (or more specifically, no-ghost conditions) vanish}, we have then performed the analysis of this same theory on a different background, a Bianchi I manifold. This background is homogeneous but not isotropic and gives a significantly better insight into the behavior of the theory just outside the isotropic limit.
In particular, one can study the behavior of the perturbations in several regions of the manifolds. One region of interest is the region of parameter space (that includes also the values of the ratio of the different scale factors and their time derivatives), which is not far away from the isotropic limit. In particular, we have found it useful to study a manifold endowed with metric $ds^2=-dt^2+a(t)^2\,dx^2+b(t)^2\,(dy^2+dz^2)$, which leaves isotropic a two-dimensional subspace. On this background, we now have the following background possibility for the vector field $A_\mu dx^\mu = A_0(t)\,dt+A_1(t)\,dx$.\footnote{The presence of $A_1(t)\neq0$ complicates the coupling among the perturbation fields.} Although the complication of the background makes its analysis harder, we found a way to shed some light on the behavior of the perturbations just away from the isotropic regime.\footnote{We studied the no-ghost condition and the discriminant equation determining the speeds of propagation on single examples of parameter space, i.e.\ after choosing some particular numerical values for the parameters of the theory and the values of $a,b,\dot{a},\dot{b}, A_0,\dot{A}_0, A_1,\dot{A}_1$. This substitution was performed a posteriori, that is after determining all the coefficients of the discriminant equations. Although the analysis is exempted from approximations (besides the standard WKB one), still we have not analyzed the whole parameter space in every detail. This means that, although we looked at several different (in our opinion) relevant examples, we might have missed some parameter space compatible with stability, but a study that requires a deep analysis of the whole parameter space, even in the fully anisotropic configuration, although interesting, is beyond the scope of this paper.}

Indeed the phenomenology of the perturbations in Bianchi I shows the presence of instabilities and strong coupling regimes for the modes in more detail, and we now summarize our findings for the even {and odd} modes. First of all, two modes propagate in the odd sector and four in the even one. This is consistent with the number of fields that were propagating on FLRW. On the other hand, we have found several new pathological behaviors which makes the theory more complicated to be understood compared to the FLRW case. In particular, one of the no-ghost conditions becomes proportional to the following quantity $(k-\mathcal{A}\,q)^2$, ($\mathcal{A}>0$), suggesting a new scale-dependent strong coupling issue. This issue was not present in FLRW. 
Second, if we look into the propagation in the $x$-direction, we find that one of the modes propagates with a modified dispersion relation of the kind $\omega^2 = \mathcal{A} k^2/a^2$ ($\mathcal{A}>0$), another mode is stable ($c_s^2=\mathcal{B}$, $\mathcal{B}>0$), whereas for the remaining mode, we find the following solutions $(c_s-\mathcal{C})^2$,  $\mathcal{C}>0$, indicating a propagation in only the positive $x$-direction. 
The meaning of these \textit{forward-propagating modes} is still obscure, but probably less worrisome than the other pathologies of the theory here studied.

In particular, along the $y{-}z$-direction, in the isotropic subspace, we find that one of the even modes still propagates with zero speed of propagation, $c_s^2=0$. Another mode is stable ($c_s^2=\mathcal{D}$, $\mathcal{D}>0$), and, finally, the remaining two modes are unstable, that is for them we have $c_s^4+\mathcal{E}c_s^2+\mathcal{F}=0$, leading to two solutions of the kind $c_s=\mathcal{A}_1\pm i\mathcal{A}_2$, with real $\mathcal{A}_{1,2}$. This in turn leads to an exponential growth for the perturbation with the instability growing large in a short time (compared to $1/H_a$ or $1/H_b$).

This behavior makes these solutions difficult to analyze from the phenomenological aspects in cosmology. How do we get out of this impasse?\footnote{Trying to set $A_0$ to vanish, i.e.~looking for the other branch of solutions, might not be a solution after all. We find that for a small but non-zero value of $A_0$, close to the isotropic FLRW case, in Bianchi I, more instabilities tend to appear. In particular, one extra mode becomes unstable when propagating on the $x$-direction or in the $y{-}z$-isotropic plane. If the $A_0(t)=0$ branch is believed to be stable, then it would be an isolated point-size solution in the whole parameter space of $A_0(t)$.} It is not clear to point to a safe direction for these theories. It is possible that another background could be stable. This would be an interesting finding to study the behavior of the theory in a controlled way. However, such knowledge is missing at the moment. It is also not clear how this background might save the cosmology for these models. Going outside the FLRW isotropic limit, made the perturbations not only have some strong coupling issues (on more, compared to the issue of $c_s^2=0$ that was plaguing the FLRW solutions) but also develop a short-time scale instability. Therefore, reducing the isotropy has led to the flourishing of extra ($k/q$-dependent) strong coupling issues (besides the one already present in the FLRW case, namely $c_s^2=0$) and Laplacian instabilities. Another point to discuss is the instability. This entire picture suggests that the theory wants to choose a different, anisotropic, and possibly inhomogeneous background, or, because of the two non-trivial strong-coupling issues, the theory has no predictive power on these backgrounds. Fixing both these issues at the moment does not look like an easy task, but if the theory has something to do with our world, then we need to find some way to solve the problems highlighted here.

{Overall, by studying the propagating modes of the Proca theory with non-minimal coupling to gravity, we have found surprising results. While we have confirmed that the theory propagates an additional scalar dof, we have found that the speed of propagation of this mode is zero, in a homogeneous and isotropic background. While we have kept the analysis only for the linearized theory, this suggests that the non-linear terms could be important as well. In order to investigate this further, we have extended our analysis also to the Bianchi I background. There, the speed of propagation of the modes still exists, but is now appearing only in the isotropic subspace, where now, in addition, one finds a tachyonic instability. The fact that the speed of propagation no longer vanishes in the $x$-direction could give us the hope that the theory could be well defined away from the isotropic space, and thus gives a possible solution to the problem of vanishing speed of propagation, in the sense that the theory could only be defined away from the purely homogeneous and isotropic case, but could maybe be realized in a well-defined way in the close limit to this approximation.  If that is the case, one should keep in mind another issue -- the rise of instabilities, whose number might even increase with an increase of anisotropy. 

While this seems to be an issue at this stage, we should take into account the possibility that these instabilities might not be too important. This is due to another type of strong coupling, that we have found in the homogeneous and isotropic case. Namely, we have found that in the limit when either the background solution of the temporal component vanishes, the coupling constants vanish, or the coupling takes the form of the Einstein tensor, the kinetic term for the additional scalar dof vanishes. At the linearized order, this would seem like a discontinuity -- after all, the theories where the previous limits take exact equalities describe only three scalar dof. Therefore, following the intuition on other gauge theories, one could expect that a mechanism analogous to the Vainshtein mechanism might take place here as well, and possibly hide this issue. In addition, we can notice that the instabilities are absent in the frame in which spatial isotropy is broken for some choices of parameters. Therefore, it may also happen that once one removes the isotropy of the subspace, the theory will be well-behaved, with a modified dispersion relation. To investigate either of the two possibilities one needs to study the more general background.  However, Proca theory with non-minimal coupling to gravity in the Jordan frame presents without a doubt a no-easy challenge that should be tackled in future studies.
}

\begin{center}
    \textbf{\textsc{Acknowledgments}}\\
\end{center}
\textit{A. H.\ and ADF would like to thank Misao Sasaki for illuminating discussions, especially for the background solutions. A. H.\ would also like to thank Elisa G. M. Ferreira, Jinn-Ouk Gong, Seong Chang Park, Yuichiro Tada,  Shuichiro Yokoyama and Ying Li Zhang for useful discussions. The work of A. H.\ was supported by the World Premier International Research Center Initiative (WPI), MEXT, Japan. The work of ADF was supported by the JSPS Grants-in-Aid for Scientific Research No.~20K03969.}

\section{Appendix}
\subsection{Action coefficients in the FLRW Universe}
\subsubsection{Coefficients associated with (\ref{FRWStartingL})}\label{APP1.1}
\begin{equation}
    \begin{split}
        c_1(t)=&-3 a^{3} \left(\left(\beta_{1}+\beta_{2}\right) A_{0}^{2}+\Mpl^{2}\right) \qquad\qquad \qquad\qquad
 c_2(t)=3\left(2 \beta_{1}+ \beta_{2}\right) a^{3} A_{0}^{2}\\
 c_3(t)=&-3\left(2 \beta_{1}+\beta_{2}\right) a^{3} A_{0}  \qquad\qquad \qquad\qquad \qquad\quad\;
 c_4(t)=-\frac{a}{2}\\
 c_5(t)=&-3\left(2 \beta_{1}+\beta_{2}\right)  a^{3} A_{0} H  \qquad\qquad\qquad\qquad\qquad\;
c_6(t)=-9 a^{3} H \left(\left(\beta_{1}+\frac{\beta_{2}}{2}\right) A_{0}^{2}+\Mpl^{2}\right)\\
c_7(t)=&-3 a^{2} \left(3 a H \left(\beta_{1}-\frac{\beta_{2}}{2}\right) A_{0}^{2}-6 A_{0} \dot{A}_0 \left(\beta_{1}+\frac{\beta_{2}}{2}\right) a +a H \,\Mpl^{2}\right)\\
c_8(t)=&-6 a^{2} \left(\dot{A}_0 \left(\beta_{1}+\frac{\beta_{2}}{2}\right) a -A_{0} a H \left(\beta_{1}-\frac{\beta_{2}}{2}\right)\right)\\
c_9(t)=&2 \beta_{2} A_{0} a  \qquad\qquad \qquad\qquad \qquad\qquad \qquad\quad\quad\;
c_{10}(t)=a\\
c_{11}(t)=&-a \left(A_{0}^{2} \beta_{1}+\Mpl^{2}\right)  \qquad\qquad\qquad\qquad \qquad\quad\;\;
c_{12}(t)=-\frac{a}{2}\\
c_{13}(t)=&\frac{a}{2 }\left(m^{2} +\left(-6  H^{2}+R \right) \left(\beta_{1}+\frac{\beta_{2}}{6}\right)+6 H^{2} \left(\beta_{1}+\frac{\beta_{2}}{3}\right)\right)\\
c_{14}(t)=&\left(2 \beta_{1}+\beta_{2}\right) A_{0}^{2} a  \qquad\qquad\qquad\qquad \qquad\qquad\;\;\;
c_{15}(t)=-2 a H \beta_{2} A_{0}\\
c_{16}(t)=&-4 \beta_{1} A_{0} a   \qquad\qquad\qquad\qquad \qquad\qquad\;\;\qquad\;
c_{17}(t)=-2 \left(-A_{0}^{2} \beta_{1}+\Mpl^{2}\right) a\\
c_{18}(t)=&-\left(2 \beta_{1}+\beta_{2}\right) a A_{0}  \qquad\qquad \qquad\qquad \qquad\quad\;\;
c_{19}(t)=-2 a H \beta_{2} A_{0}\\
c_{20}(t)=&-\frac{3 a^3 }{2}\left(\left(A_{0}^{2} m^{2}+2 \Lambda \,\Mpl^{2}\right)+\left(3\left( \beta_{1}+\frac{ \beta_{2}}{2}\right) A_{0}^{2}+\Mpl^{2}\right) \left(-6 H^{2}+R \right)+6  H^{2} \left(3 A_{0}^{2} \beta_{1}+\Mpl^{2}\right)\right)\\
c_{21}(t)=&3 a^3 A_{0} \left(m^{2}+\left(\beta_{1}+\frac{\beta_{2}}{2}\right) \left(-6 H^{2}+R \right)+6 \beta_{1}  H^{2}\right)\\
c_{22}(t)=&- A_{0} a^3\left(m^{2} +3 \left(\beta_{1}+\frac{\beta_{2}}{2}\right) \left(-6 H^{2}+R \right)+18 \beta_{1}  H^{2}\right)\\
c_{23}(t)=&\frac{3a}{4} \left(m^{2} a^{2}A_0^2+\frac{10}{3}\left(\beta_{1}+\frac{\beta_{2}}{2}\right) \left(-6 a^{2} H^{2}+R \,a^{2}\right)A_0^2+10 a^{2} H^{2} \left(\beta_{1}-\beta_{2}\right)A_{0}^{2}\right.\\&\left.-20 \dot{A}_0 A_{0} a^{2} \left(\beta_{1}+\frac{\beta_{2}}{2}\right) H +\frac{2 }{3}\Mpl^{2} \left(-9 a^{2} H^{2}+\Lambda \,a^{2}\right)\right)\\
 c_{24}(t)=&\frac{3 a }{4}\left(\left(A_{0}^{2} m^{2}+4 \Ddot{A}_0 \left(\beta_{1}+\frac{\beta_{2}}{2}\right) A_{0}+\left(4 \beta_{1}+2 \beta_{2}\right) \dot{A}_0^{2}-2 \Lambda \,\Mpl^{2}\right) a^{2}+8 a^2 H \dot{A}_0 \left(\beta_{1}+\beta_{2}\right) A_{0}\right.\\&\left.+
 \frac{2}{3} \left(-6 a^{2} H^{2}+R \,a^{2}\right) \left(\beta_{1}+\beta_{2}\right) A_{0}^{2}+\frac{2}{3} \left(-6 a^{2} H^{2}+R \,a^{2}\right) \Mpl^{2} +2 \left(\left(\beta_{1}+\beta_{2}\right) A_{0}^{2}+\Mpl^{2}\right) a^{2} H^{2}\right)\\
c_{25}(t)=&\frac{a}{2}\left(m^{2} a^{2}+\left(\beta_{1}+\frac{\beta_{2}}{2}\right) \left(-6 a^{2} H^{2}+R \,a^{2}\right)+6 \beta_{1} a^{2} H^{2}\right) 
    \end{split}
\end{equation}

\subsubsection{Coefficients associated with (\ref{FRWF1})}\label{APP1.2}

\begin{equation}
    \begin{split}
        d_1(t)=&-3 a^{3} \left(\left(\beta_{1}+\beta_{2}\right) A_{0}^{2}+\Mpl^{2}\right) \qquad \qquad\qquad \qquad
        d_2(t)=\left(3 \beta_{1}+ \beta_{2}\right) a^{3} A_{0}^{2}\\
        d_3(t)=&-\left(3 \beta_{1}+ \beta_{2}\right) a^{3} A_{0} \qquad\qquad \qquad\qquad \quad\;\;\;\;\qquad
        d_4(t)=\frac{a \,k^{2}}{2}\\
        d_5(t)=&-\left(3 \beta_{1}+ \beta_{2}\right) a^{3} A_{0} H\qquad\qquad \qquad\qquad \quad\;\qquad
        d_6(t)=-\frac{9}{2}a^{3} H \left(\left(\beta_{1}+\frac{\beta_{2}}{2}\right) A_{0}^{2}+\Mpl^{2}\right)\\
        d_7(t)=&-\frac{3}{2}a^{2} \left(3 a H \left(\beta_{1}-\frac{\beta_{2}}{2}\right) A_{0}^{2}-6 A_{0} \dot{A}_0 \left(\beta_{1}+\frac{\beta_{2}}{2}\right) a +a H \,\Mpl^{2}\right)\\
        d_8(t)=&-3 a^{2} \left(\dot{A}_0 \left(\beta_{1}+\frac{\beta_{2}}{2}\right) a -A_{0} a H \left(\beta_{1}-\frac{\beta_{2}}{2}\right)\right)\\
        d_9(t)=&- a A_{0} k^{2} \beta_{2}\qquad\qquad \qquad\qquad \qquad\qquad \qquad\quad
        d_{10}(t)=-\frac{1}{2}a \,k^{2}\\
        d_{11}(t)=&a \left(A_{0}^{2} \beta_{1}+\Mpl^{2}\right) k^{2}\qquad\qquad \qquad\qquad \qquad\qquad 
        d_{12}(t)=\frac{a \,k^{2}}{2}\\
        d_{13}(t)=&-\frac{\beta_{2} \left(\left(12 \beta_{1}+3 \beta_{2}\right) a^{2} H^{2}+m^{2} a^{2}\right) k^{2}}{\left(6 \beta_{1}+3 \beta_{2}\right) a}\\
        d_{14}(t)=&-2 a \left(\left(\left(3 \beta_{1}+3 \beta_{2}\right) a^{2} H^{2}+\frac{3 m^{2} a^{2}}{4}+k^{2} \left(\beta_{1}+\frac{\beta_{2}}{2}\right)\right) A_{0}^{2}+6 \dot{A}_0 A_{0} a^{2} \left(\beta_{1}+\frac{\beta_{2}}{2}\right) H +\frac{3 \Mpl^{2} a^{2} H^{2}}{2}\right)\\
        d_{15}(t)=& \beta_{2} A_{0} k^{2} a H\qquad\qquad \qquad\qquad \qquad\qquad \qquad\quad
        d_{16}(t)=2 a A_{0} \beta_{1} k^{2}\\
        d_{17}(t)=&\frac{2 a}{2 \beta_{1}+\beta_{2}} \left(-\left(\beta_{1}+\frac{\beta_{2}}{2}\right) \left(\frac{9 \left(\beta_{1}+\beta_{2}\right) a^{2} H^{2}}{2}-\frac{3 m^{2} a^{2}}{4}+\beta_{1} k^{2}\right) A_{0}^{2}\right.\\&\left.-9 \left(\beta_{1}+\frac{\beta_{2}}{2}\right)^{2} a^{2} H \dot{A}_0 A_{0}+\Mpl^{2} \left(\frac{9 \left(-\beta_{1}-\beta_{2}\right) a^{2} H^{2}}{2}+\frac{3 m^{2} a^{2}}{4}+k^{2} \left(\beta_{1}+\frac{\beta_{2}}{2}\right)\right)\right)\\
        d_{18}(t)=& A_{0} a \left(m^{2} a^{2}+k^{2} \left(\beta_{1}+\frac{\beta_{2}}{2}\right)\right)
    \end{split}
\end{equation}

\subsubsection{Coefficients associated with (\ref{FRWF2})}\label{APP1.3}
Below, we will present most of the coefficients for the action, which the reader can use as the check. However, due to its length, and because it is almost impossible to read and will not bring anything interesting,  we will omit the coefficient $e_4$. 
\begin{equation}
    \begin{split}
        e_3=&\frac{k \,a^{2}\dot{A}_0 }{2\left(\left(\beta_{1}+\beta_{2}\right) a H A_{0}^{2}+\dot{A}_0 A_{0} \left(\beta_{1}+\frac{\beta_{2}}{2}\right) a +a H \,\Mpl^{2}\right)^{2}}\times \left(2 \left(\beta_{1}+\frac{\beta_{2}}{2}+\frac{3}{4}\right) a H \beta_{2} \left(\beta_{1}+\beta_{2}\right) A_{0}^{4}\right.\\&\left.+2 \left(\beta_{1}+\frac{\beta_{2}}{2}\right) \left(\frac{\beta_{2}^{2}}{2}+\left(\beta_{1}+\frac{3}{2}\right) \beta_{2}+\frac{3 \beta_{1}}{2}\right) a \dot{A}_0 A_{0}^{3}+2 \left(\frac{\beta_{2}^{2}}{2}+\left(\beta_{1}+\frac{5}{4}\right) \beta_{2}+\frac{\beta_{1}}{2}\right) a H \,\Mpl^{2} A_{0}^{2}\right.\\&\left.+\Mpl^{2} \dot{A}_0 a \left(\beta_{1}+\frac{\beta_{2}}{2}\right) A_{0}+\Mpl^{4} a H \right) 
    \end{split}
\end{equation}
\begin{equation}
    \begin{split}
         e_5=-\frac{\left(\left(12 \beta_{1}+3 \beta_{2}\right) a^{2} H^{2}+a^{2} m^{2}\right) \beta_{2}}{\left(6 \beta_{1}+3 \beta_{2}\right) a}
    \end{split}
\end{equation}

\begin{equation}
    \begin{split}
        e_6 &= \frac{2 k A_{0} \beta_{2}}{\left(\beta_{1} + \frac{\beta_{2}}{2}\right) \left(\left(\beta_{1} + \beta_{2}\right) a H A_{0}^{2} + \dot{A}_0 A_{0} \left(\beta_{1} + \frac{\beta_{2}}{2}\right) a + a H \, \Mpl^{2} \right)^2} \\
        & \times \left( - \frac{\left( a^{2} m^{2} + 12 a^{2} H^{2} \left(\frac{\beta_{2}}{4} + \beta_{1}\right) \right) \left(\beta_{1} + \beta_{2}\right)^2 a H A_{0}^{4}}{6} \right. \\
        & \quad \left. - \left( \beta_{1} + \beta_{2} \right) \left( \frac{\dot{A}_0 m^{2} a^{2}}{6} + \left( -a H \ddot{A}_0 + \frac{\dot{A}_0 \left( -6 a^{2} H^{2} + R a^{2} \right)}{6 a} \right) \left(\beta_{1} + \frac{\beta_{2}}{2}\right) a \right. \right. \\
        & \quad \left. \left. - \frac{\dot{A}_0 \beta_{2} a^{2} H^{2}}{2} \right) \left(\beta_{1} + \frac{\beta_{2}}{2}\right) a A_{0}^{3} \right. \\
        & \quad \left. + a H \left( \left( \left( \beta_{1} + \frac{\beta_{2}}{2} \right)^3 \dot{A}_0^{2} - \frac{m^{2} \Mpl^{2} \left(\beta_{1} + \beta_{2}\right)}{3} \right) a^{2} - 4 \left( \beta_{1} + \beta_{2} \right) a^{2} H^{2} \left( \frac{\beta_{2}}{4} + \beta_{1} \right) \Mpl^{2} \right) A_{0}^{2} \right. \\
        & \quad \left. + \left( \left( \left( \beta_{1} + \frac{\beta_{2}}{2} \right)^2 \dot{A}_0^{2} - \frac{m^{2} \Mpl^{2}}{6} \right) \dot{A}_0 a^{2} \right. \right. \\
        & \quad \left. \left. - \left( -a H \ddot{A}_0 + \frac{\dot{A}_0 \left( -6 a^{2} H^{2} + R a^{2} \right)}{6 a} \right) \left(\beta_{1} + \frac{\beta_{2}}{2}\right) \Mpl^{2} a + \frac{\dot{A}_0 \Mpl^{2} \beta_{2} a^{2} H^{2}}{2} \right) \left(\beta_{1} + \frac{\beta_{2}}{2}\right) a A_{0} \right. \\
        & \quad \left. + 2 a H \left( \left( \left( \beta_{1} + \frac{\beta_{2}}{2} \right)^2 \dot{A}_0^{2} - \frac{m^{2} \Mpl^{2}}{12} \right) a^{2} - a^{2} H^{2} \left( \frac{\beta_{2}}{4} + \beta_{1} \right) \Mpl^{2} \right) \Mpl^{2} \right)
    \end{split}
\end{equation}
\subsubsection{Coefficients associated with (\ref{FRWF3})}\label{APP1.4}

\begin{equation}
    \begin{split}
        &v_1(t)=\frac{a^{3}}{2}\\
        &v_2(t)=\frac{a}{2}\\
        &v_3(t)=-\frac{1}{2}\left(\left(m^{2} a^{2}+\left(\beta_{1}+\frac{\beta_{2}}{6}+\frac{1}{6}\right) \left(-6 a^{2} H^{2}+R \,a^{2}\right)+6 a^{2} H^{2} \left(\beta_{1}+\frac{\beta_{2}}{3}+\frac{1}{6}\right)\right) a\right)\\
        &v_4(t)=-\frac{\left(\left(\beta_{1}+\beta_{2}\right) A_{0}^{2}+\Mpl^{2}\right) a}{4}\\
        &v_5(t)=-\frac{a}{4}\left(\left(m^{2} a^{2}+2 \left(-6 a^{2} H^{2}+R \,a^{2}\right) \left(\beta_{1}+\frac{\beta_{2}}{2}\right)+6 a^{2} H^{2} \left(\beta_{1}-\beta_{2}\right)\right) A_{0}^{2}\right.\\&\left.-12 \left(\beta_{1}+\frac{\beta_{2}}{2}\right) A_{0} a^{2}\dot{A}_0 H +2 \Mpl^{2} \left(-3 a^{2} H^{2}+\Lambda \,a^{2}\right)\right)\\
        &v_6(t)=-\frac{A_{0} \beta_{2} a}{2}
    \end{split}
\end{equation}

\subsection{Fixed point analysis for the background equations.}

In section 3 we have studied the fixed point analysis for the background solutions in the FLRW universe for specific choices of couplings. Here, we will generalize this analysis to other values as well.

The equations (\ref{dotH}) and (\ref{dottemp}) make an autonomous system of equations with which we will be able to determine the evolution of the Hubble parameter and the temporal component. It can be easily checked that the equation (\ref{acceqFLRW2}) is automatically satisfied if we substitute the two. 

Let us now consider the fixed points. In order to work with the dimensionless variables, we will set $\Mpl=1$. The fixed points are defined by: 
\begin{equation}
    \dot{A}_0=0\qquad \text{and}\qquad \dot{H}=0. 
\end{equation}
This leads us to four fixed points, given in the form $(H, A_0)$: 
\begin{equation}
    \begin{split}
        \left(\pm\frac{m}{\sqrt{-12 \beta_{1}-3 \beta_{2}}}, \pm\frac{2 \sqrt{\left(\Lambda \left(4 \beta_{1}+\beta_{2}\right)+m^{2}\right) \left(\beta_{1}-\frac{\beta_{2}}{2}\right)}}{\left(2 \beta_{1}-\beta_{2}\right) m}\right),
    \end{split}
\end{equation}
%\ADF{which are real provided the radicand above, $\left[\Lambda \left(4 \beta_{1}+\beta_{2}\right)+m^{2}\right] \left(\beta_{1}-\frac{\beta_{2}}{2}\right)$, is non-negative.}
which are real provided the radicand above, $\left[\Lambda \left(4 \beta_{1}+\beta_{2}\right)+m^{2}\right] \left(\beta_{1}-\frac{\beta_{2}}{2}\right)$, is non-negative.
We will symbolically refer to them as $(\pm, \pm)$. They take a particularly simple form in the absence of the cosmological constant: 
\begin{equation}
    \begin{split}
        \left(\pm\frac{m}{\sqrt{-12 \beta_{1}-3 \beta_{2}}}, \pm\frac{2}{\sqrt{4 \beta_{1}-2 \beta_{2}}} \right)
    \end{split}
\end{equation}
Clearly, the nature of the fixed points will depend on the parameters of the theory.  In order to study their stability, we compute the Jacobian matrix, 
\begin{equation}
    J=\begin{bmatrix}
f_{,H} & f_{,A_0} \\
g_{,H} & g_{,A_0} 
\end{bmatrix}
\end{equation}
where 
\begin{equation}
\begin{split}
        f(t) &= 
\frac{\left(-12 \beta_{1}-3 \beta_{2}\right)H^{2}-m^{2}}{6 \beta_{1}+3 \beta_{2}}\qquad\text{and}\qquad \\ g(t) &= 
\frac{\left(-6+\left(-6 \beta_{1}-6 \beta_{2}\right) A_{0}^{2}\right)H^{2}-A_{0}^{2} m^{2}+2 \Lambda}{\left(6 \beta_{2}+12 \beta_{1}\right) A_{0}H}.
\end{split}
\end{equation}
The eigenvalues are given by: 
\begin{equation}
    \begin{split}
        \lambda_1&= -\frac{2H \left(4 \beta_{1}+\beta_{2}\right)}{\beta_{2}+2 \beta_{1}} \qquad\text{and}\\
        \lambda_2&= -\frac{6H^{2} A_0^{2} \beta_{1}+6H^{2} A_0^{2} \beta_{2}+A_0^{2} m^{2}-6H^{2}+2 \Lambda}{6 \left(\beta_{2}+2 \beta_{1}\right)H A_0^{2}}.
    \end{split}
\end{equation}
The case for vanishing cosmological constant corresponds to setting $\Lambda=0$ in the previous expression. Let us now substitute the fixed points in the previous expression. For $(+, \pm)$, we find: 
\begin{equation}
    \begin{split}
        \lambda_1&= -\frac{2 m \left(4 \beta_{1}+\beta_{2}\right)}{\sqrt{-12 \beta_{1}-3 \beta_{2}}\, \left(\beta_{2}+2 \beta_{1}\right)}  \qquad\text{and}\qquad 
        \lambda_2= -\frac{m \sqrt{-12 \beta_{1}-3 \beta_{2}}\, \left(2 \beta_{1}-\beta_{2}\right)}{3 \left(4 \beta_{1}+\beta_{2}\right) \left(\beta_{2}+2 \beta_{1}\right)} .
    \end{split}
\end{equation}
For $(-, \pm)$, we find: 
\begin{equation}
    \begin{split}
        \lambda_1&= \frac{2 m \left(4 \beta_{1}+\beta_{2}\right)}{\sqrt{-12 \beta_{1}-3 \beta_{2}}\, \left(\beta_{2}+2 \beta_{1}\right)}  \qquad\text{and}\qquad 
        \lambda_2= \frac{m \sqrt{-12 \beta_{1}-3 \beta_{2}}\, \left(2 \beta_{1}-\beta_{2}\right)}{3 \left(4 \beta_{1}+\beta_{2}\right) \left(\beta_{2}+2 \beta_{1}\right)} .
    \end{split}
\end{equation}
Curiously, we can notice that the cosmological constant has dropped out of the previous expressions. They match also the case when $\Lambda=0$. This means that the cosmological constant will not play a role in the stability of the background solutions for the points that we have found in the above. It will, however, contribute to the no-ghost conditions, and thus play the role in the stability analysis of the degrees of freedom. 

We can notice that if $4\beta_1+\beta_2>0$, the previous eigenvalues become imaginary for all four fixed points. In this case, the points are then elliptic. 

Let us now assume that $4\beta_1+\beta_2<0$. Then, we find: 
\begin{enumerate}[label=(\roman*)]
    \item The points $(+,\pm)$ are stable, if $\beta_1>0$.   
    \item If $\beta_1<0$, and $-2\beta_1>\beta_2>2\beta_1$, the point $(+,\pm)$ is a saddle point. 
    \item  If $\beta_1<0$, and $2\beta_1>\beta_2$, the point $(+,\pm)$ is stable. 
    \item For $\beta_1>0$, and $\beta_2<-4\beta_1$, the point $(-,\pm)$ is unstable. 
    \item For $\beta_1<0$, and $\beta_2<2\beta_1$, the point $(-,\pm)$ is unstable. 
    \item For $\beta_1<0$, and $-4\beta_1>\beta_2>2\beta_1$, the point $(-,\pm)$ is a saddle point. 
\end{enumerate}

We can notice that the cosmological constant has no effect on the expansion rate of the universe, or on the stability of the fixed points. It only shifts the solution for the $A_0$ component. However, while we were studying the degrees of freedom, we have substituted the constraint equation for the cosmological constant. This has led us to the following no-ghost conditions: 
\begin{equation}
      \left(\beta_{1}+\beta_{2}\right) A_{0}^{2}+\Mpl^{2}>0
\end{equation} 
Therefore, while the cosmological constant does not have a big influence on the background evolution, it affects the no-ghost conditions.

\subsection{Coefficients associated with the odd modes}\label{APPodd}
\begin{equation}
    \begin{split}
        k_1(t)=&-\frac{a}{2}\\
         k_2(t)=&-\frac{\left(\left(A_{0}^{2} \beta_{1}+\frac{1}{2} A_{0}^{2} \beta_{2}+\Mpl^{2}\right) a^{2}-\left(\beta_{1}+\frac{\beta_{2}}{2}\right) A_{1}^{2}\right) }{a^{3}}\\
        k_3(t)=&\frac{\left(-3 A_{0}^{2} \beta_{1}-A_{0}^{2} \beta_{2}-3 \Mpl^{2}\right) a^{2}+3 A_{1}^{2} \beta_{1}}{4 a \,b^{2}}\\
        k_4(t)=&-\frac{a}{2 b^{2}}\\
        k_5(t)=&\frac{2}{ \,a^{2}}  \frac{3}{4}  \left(\left(\left(\beta_{1}+\frac{\beta_{2}}{2}\right) A_{0}^{2}+\frac{\Mpl^{2}}{3}\right) a^{2}-\frac{\left(\beta_{1}+\frac{\beta_{2}}{2}\right) A_{1}^{2}}{3}\right) \ddot{a}\\&+ \frac{2}{b \,a^{2}} \left(\frac{a}{2}\left(\left(3 \left(\beta_{1}+\frac{\beta_{2}}{2}\right) A_{0}^{2}+\Mpl^{2}\right) a^{2}-A_{1}^{2} \beta_{1}\right) \ddot{b}\right)\\&+\left(\left(H_{b} \left(H_{a}-\frac{H_{b}}{4}\right) \beta_{1}-\frac{H_{a} H_{b} \beta_{2}}{4}-\frac{H_{b}^{2} \beta_{2}}{2}+\frac{m^{2}}{8}\right) A_{0}^{2}-\frac{3 \left(H_{b}^{2}-\frac{\Lambda}{3}\right) \Mpl^{2}}{4}\right) 2a\\&+\left(\frac{3 H_{b}^{2} \beta_{1} A_{1}^{2}}{4}+\frac{m^{2} A_{1}^{2}}{8}+\frac{\dot{A}_1^{2}}{8}\right) ba \\
k_6(t) =& -\frac{4A_0}{b} \left( \frac{\ddot{a} b \left(\beta_{1}+\frac{\beta_{2}}{2}\right)}{2} 
    + a \left( \left(\beta_{1}+\frac{\beta_{2}}{2}\right) \ddot{b} 
    + \frac{b \left(\left(4 H_{a} H_{b}+2 H_{b}^{2}\right) \beta_{1}+m^{2}\right)}{4} \right) \right)
\\ 
k_7(t)=&\frac{4 a \left(\beta_{1}+\frac{\beta_{2}}{4}\right) \ddot{b}+\left(2 \ddot{a} \beta_{1}+a \left(\left(4 H_{a} H_{b}+2 H_{b}^{2}\right) \beta_{1}+H_{a} H_{b} \beta_{2}+H_{b}^{2} \beta_{2}+m^{2}\right)\right) b}{2 b}\\
k_8(t)=&\frac{\left(A_{0}^{2} \beta_{1}+\frac{1}{2} A_{0}^{2} \beta_{2}+\Mpl^{2}\right) \left(H_{a}+2 H_{b}\right) a^{2}-\left(2 H_{b} \beta_{1}+H_{a} \left(\beta_{1}+\frac{\beta_{2}}{2}\right)\right) A_{1}^{2}}{a}\\
k_9(t)=&\frac{2 \beta_{2} H_{b} A_{0} A_{1}}{a}\\
k_{10}(t)=&\frac{A_{1} \beta_{2} \left(H_{a}-2 H_{b}\right)}{2 a}\\
k_{11}(t)=&\frac{\left(3 A_{0}^{2} \beta_{1}+A_{0}^{2} \beta_{2}+3 \Mpl^{2}\right) a^{2}-3 \left(\beta_{1}+\frac{\beta_{2}}{3}\right) A_{1}^{2}}{4 a^{3}}\\
k_{12}(t)=&-\frac{\dot{A}_1}{a}\\
k_{13}(t)=&\frac{2 A_{1} A_{0}}{b \,a^{2}} \left(\frac{b \left(2 \beta_{1}+\beta_{2}\right) \ddot{a}}{2}+a \left(\left(2 \beta_{1}+\beta_{2}\right) \ddot{b}+\frac{b \left(\left(4 H_{a} H_{b}+2 H_{b}^{2}\right) \beta_{1}+m^{2}\right)}{2}\right)\right)\\
k_{14}(t)=&-\frac{4A_1}{ba^2} \left(\frac{\ddot{a} \left(\beta_{1}+\frac{\beta_{2}}{2}\right) b}{2}\right.\\&\left.+a \left(\left(\beta_{1}+\frac{\beta_{2}}{4}\right) \ddot{b}+\frac{b \left(\left(4 H_{a} H_{b}+2 H_{b}^{2}\right) \beta_{1}+H_{b} \left(H_{a}+H_{b}\right) \beta_{2}+m^{2}\right)}{4}\right)\right) 
   \end{split}
\end{equation}
\subsection{Coefficients associated with (\ref{bianchiLProca})}\label{APP2}
\begin{equation*}
    \begin{split}
&f_1(t)=-\frac{b^{2} A_{1}^{2} \left(2 \beta_{1}+\beta_{2}\right)}{4 a}\\
&f_2(t)=-\frac{a \,b^{2} \left(2 \beta_{1}+\beta_{2}\right) A_{0}^{2}}{2}\\
&f_3(t)=-\frac{a \,b^{2} \left(2 \beta_{1}+\beta_{2}\right) A_{0}^{2}}{2}\\
&f_4(t)=\frac{b^{2} A_{1}^{2} \left(2 \beta_{1}+\beta_{2}\right)}{2 a}\\
&f_5(t)=-\frac{b^{2} \left(2 \beta_{1}+\beta_{2}\right) A_{0} A_{1}}{2 a}\\
&f_6(t)=-\frac{b^{2} A_{1}^{2}}{2 a}\\
&f_7(t)=-\frac{b^{2} A_{1}^{2}}{2 a}\\
&f_8(t)=\frac{\left(2 A_{0}^{2} \beta_{1}+2 A_{0}^{2} \beta_{2}+2 \Mpl^{2}\right) a^{2}-2 \left(\beta_{1}+\frac{\beta_{2}}{2}\right) A_{1}^{2}}{4 a}\\
&f_9(t)=-\frac{a A_{0} \beta_{2}}{2}\\
&f_{10}(t)=\frac{\left(-2 A_{0}^{2} \beta_{1}-A_{0}^{2} \beta_{2}-2 \Mpl^{2}\right) a^{2}+2 A_{1}^{2} \beta_{1}}{2 a}\\
&f_{11}(t)=-\frac{b^{2} A_{1} A_{0}}{a}\\
&f_{12}(t)=-a A_{0}\\
&f_{13}(t)=-\frac{b^{2} \dot{A}_1 A_{1}}{a}\\
&f_{14}(t)=-\frac{b^{2} \left(2 \beta_{1}+\beta_{2}\right) A_{0}^{2}}{a}\\
&f_{15}(t)=-\frac{\left(\left(A_{0}^{2} \beta_{1}+\frac{1}{2} A_{0}^{2} \beta_{2}+\Mpl^{2}\right) a^{2}-\left(\beta_{1}+\frac{\beta_{2}}{2}\right) A_{1}^{2}\right) b^{2}}{a^{3}}\\
&f_{16}(t)=\frac{\left(\left(A_{0}^{2} \beta_{1}+\frac{1}{2} A_{0}^{2} \beta_{2}+\Mpl^{2}\right) \left(H_{a}+2 H_{b}\right) a^{2}-\left(2 H_{b} \beta_{1}+H_{a} \left(\beta_{1}+\frac{\beta_{2}}{2}\right)\right) A_{1}^{2}\right) b^{2}}{a^{3}}\\
&f_{17}(t)=-\frac{\left(\left(3 A_{0}^{2} \beta_{1}+\frac{3}{2} A_{0}^{2} \beta_{2}+\Mpl^{2}\right) a^{2}-\left(\beta_{1}+\frac{\beta_{2}}{2}\right) A_{1}^{2}\right) b^{2}}{a^{3}}\\
&f_{18}(t)=-\frac{b^{2} \left(2 \beta_{1}+\beta_{2}\right) A_{1}^{2}}{2 a^{3}}\\
&f_{19}(t)=\frac{\left(A_{0}^{2} \beta_{1}+\frac{1}{2} A_{0}^{2} \beta_{2}+\Mpl^{2}\right) \left(H_{a}+2 H_{b}\right) a^{2}-\left(2 H_{b} \beta_{1}+H_{a} \left(\beta_{1}+\frac{\beta_{2}}{2}\right)\right) A_{1}^{2}}{a}\\
&f_{20}(t)=-a \left(2 \beta_{1}+\beta_{2}\right) A_{0}^{2}\\
&f_{21}(t)=\frac{\left(-6 A_{0}^{2} \beta_{1}-3 A_{0}^{2} \beta_{2}-2 \Mpl^{2}\right) a^{2}+2 A_{1}^{2} \beta_{1}}{2 a}\\
&f_{22}(t)= -\frac{A_{1}^{2} \left(4 \beta_{1}+\beta_{2}\right)}{4 a}\\
&f_{23}(t)=b^{2} a \left(H_{a}+2 H_{b}\right) \left(2 \beta_{1}+\beta_{2}\right) A_{0}^{2}\\
&f_{24}(t)=A_{0} a \left(\left(\left(\frac{H_{a}}{2}-H_{b}\right) \beta_{2}+H_{a} \beta_{1}\right) A_{0}-2 \left(\beta_{1}+\frac{\beta_{2}}{2}\right)\dot{A}_0\right) b^{2}\\
&f_{25}(t)=\frac{\left(\left(A_{0}^{2} \beta_{1}+\frac{1}{2} A_{0}^{2} \beta_{2}+\Mpl^{2}\right) \left(H_{a}+2 H_{b}\right) a^{2}+\left(2 H_{b} \beta_{1}+H_{a} \left(\beta_{1}+\frac{\beta_{2}}{2}\right)\right) A_{1}^{2}\right) b^{2}}{2 a}
  \end{split}
\end{equation*}
\begin{equation*}
    \begin{split}
&f_{26}(t)= \frac{3 b^{2}a}{2}\left(\frac{A_{0} \left(A_{0} H_{a}-2 A_{0} H_{b}-2\dot{A}_0\right) \beta_{2}}{2}+\left(A_{0}^{2} \beta_{1}+\frac{\Mpl^{2}}{3}\right) H_{a}-2 \beta_{1} A_{0}\dot{A}_0\right)\\&-\frac{3 b^{2}}{2a}\left(\beta_{1}+\frac{\beta_{2}}{2}\right) A_{1} \left(A_{1} H_{a}-\frac{2 \dot{A}_1}{3}\right)\\
&f_{27}(t)=\frac{2 A_{1}^{2} \left(\left(H_{a}+2 H_{b}\right) \beta_{1}+\frac{H_{a} \beta_{2}}{2}\right) b^{2}}{a}\\
&f_{28}(t)=-\frac{b^{2} A_{1} \left(H_{a}+2 H_{b}\right) \left(2 \beta_{1}+\beta_{2}\right) A_{0}}{a}\\
&f_{29}(t)=-\frac{3 b^{2} \left(\left(\left(\left(H_{a}-\frac{2 H_{b}}{3}\right) \beta_{2}+2 H_{a} \beta_{1}\right) A_{0}-\frac{\dot{A}_0 \left(2 \beta_{1}+\beta_{2}\right)}{3}\right) A_{1}-\frac{A_{0} \dot{A}_1 \left(2 \beta_{1}+\beta_{2}\right)}{3}\right)}{2 a}\\
&f_{30}(t)=\frac{b^{2} \left(3 A_{1} H_{a}-2 \dot{A}_1\right) A_{1} \left(2 \beta_{1}+\beta_{2}\right)}{2 a}\\
&f_{31}(t)=\frac{b^{2} \dot{A}_1 A_{1}}{a}\\
&f_{32}(t)=-\frac{b^{2} \dot{A}_1 A_{1}}{2 a}\\
&f_{33}(t)=\frac{2 \left(\frac{b A_{1}^{2} \left(2 \beta_{1}+\beta_{2}\right) \ddot{a}}{2}+\left(2 \ddot{b} A_{1}^{2} \beta_{1}+\left(\left(\left(2 H_{a} H_{b}+H_{b}^{2}\right) \beta_{1}+H_{a} H_{b} \beta_{2}-\frac{m^{2}}{2}\right) A_{1}^{2}-\frac{\dot{A}_1^{2}}{2}\right) b \right) a \right) b}{a^{2}}\\
&f_{34}(t)=\frac{b^{2} \left(2 A_{1} H_{b} \beta_{2}-\dot{A}_1\right) A_{0}}{a}\\
&f_{35}(t)=\frac{b^{2} \dot{A}_1 A_{0}}{2 a}\\
&f_{36}(t)=-\frac{\left(\frac{b A_{1}^{2} \left(2 \beta_{1}+\beta_{2}\right) \ddot{a}}{2}+a \left(2 \ddot{b} A_{1}^{2} \beta_{1}+\left(\left(\left(2 H_{a} H_{b}+H_{b}^{2}\right) \beta_{1}+H_{a} H_{b} \beta_{2}+\frac{m^{2}}{2}\right) A_{1}^{2}-\frac{\dot{A}_1^{2}}{2}\right) b \right)\right) b}{a^{2}}\\
&f_{37}(t)=\frac{A_{0} \left(\frac{b \left(2 \beta_{1}+\beta_{2}\right) \ddot{a}}{2}+\left(\left(2 \beta_{1}+\beta_{2}\right) \ddot{b}+\frac{b \left(\left(4 H_{a} H_{b}+2 H_{b}^{2}\right) \beta_{1}+m^{2}\right)}{2}\right) a \right) b A_{1}}{a^{2}}\\
&f_{38}(t)=\frac{b^{2} A_{1} A_{0} H_{b} \beta_{2}}{a}\\
&f_{39}(t)=\frac{1}{2 a}\\
&f_{40}(t)=\frac{2 A_{0} \left(\frac{b \left(2 \beta_{1}+\beta_{2}\right) \ddot{a}}{2}+\left(\left(2 \beta_{1}+\beta_{2}\right) \ddot{b}+\frac{b \left(\left(4 H_{a} H_{b}+2 H_{b}^{2}\right) \beta_{1}+m^{2}\right)}{2}\right) a \right) b A_{1}}{a^{2}}\\
&f_{41}(t)=-\frac{b^{2} A_{0}^{2}}{2 a}\\
&f_{42}(t)=-\frac{b^{2} \dot{A}_1 A_{0}}{a}\\
&f_{43}(t)=\frac{b}{a^2}\left(\frac{b A_{1}^{2} \left(2 \beta_{1}+\beta_{2}\right) \ddot{a}}{2}+a \left(2 \ddot{b} A_{1}^{2} \beta_{1}+\left(\left(\left(2 H_{a} H_{b}+H_{b}^{2}\right) \beta_{1}+H_{a} H_{b} \beta_{2}+\frac{m^{2}}{2}\right) A_{1}^{2}-\frac{\dot{A}_1^{2}}{2}\right) b \right)\right) \\
&f_{44}(t)=-\frac{4 b^{2} \beta_{1} A_{0}^{2} H_{b}}{a}\\
&f_{45}(t)=-\frac{2 A_{0} \left(\frac{b \left(2 \beta_{1}+\beta_{2}\right) \ddot{a}}{2}+\left(\left(2 \beta_{1}+\beta_{2}\right) \ddot{b}+\frac{b \left(\left(4 H_{a} H_{b}+2 H_{b}^{2}\right) \beta_{1}+m^{2}\right)}{2}\right) a \right) b A_{1}}{a^{2}}\\
 \end{split}
\end{equation*}
\begin{equation*}
    \begin{split}
    &f_{46}(t)=\frac{3 b^2}{2 a^{4}}  \left(\left(A_{0}^{2} \beta_{1}+\frac{1}{2} A_{0}^{2} \beta_{2}+\frac{1}{3} \Mpl^{2}\right) a^{2}-\left(\beta_{1}+\frac{\beta_{2}}{2}\right) A_{1}^{2}\right) \ddot{a}\\&+\frac{3 b}{2 a^{3}} \left(\left(A_{0}^{2} \beta_{2}+2 A_{0}^{2} \beta_{1}+\frac{2}{3} \Mpl^{2}\right) a^{2}-2 \left(\beta_{1}+\frac{\beta_{2}}{3}\right) A_{1}^{2}\right) \ddot{b}+ \frac{3 b^2}{2 a}A_{0}^{2} \left(\frac{2}{3} H_{a} H_{b}+H_{b}^{2}-\frac{2}{3} H_{a}^{2}\right) \beta_{1}\\&+\left(\frac{\left(-A_{0}^{2} \beta_{2}-2 \Mpl^{2}\right) H_{a}^{2}}{3}-\frac{2 H_{b} \left(A_{0}^{2} \beta_{2}+\Mpl^{2}\right) H_{a}}{3}+\frac{m^{2} A_{0}^{2}}{6}+\frac{\Mpl^{2} \left(H_{b}^{2}+\Lambda \right)}{3}\right) \frac{3 b^2}{2 a} \\&-A_{1}^{2} \frac{3 b^2}{2 a^{3}}\left(\frac{2}{3} H_{a} H_{b}+H_{b}^{2}-\frac{2}{3} H_{a}^{2}\right) \beta_{1}+\frac{3 b^2}{2 a^{3}}\frac{\left(-H_{a} H_{b} \beta_{2}-\frac{1}{2} m^{2}+H_{a}^{2} \beta_{2}\right) A_{1}^{2}}{3}+\frac{3 b^2}{2 a^{3}}\frac{\dot{A}_1^{2}}{6}\\
&f_{47}(t)=\frac{b^{2} \left(2 \beta_{1}+\beta_{2}\right) A_{0}^{2}}{a}\\
&f_{48}(t)=-\frac{2 b^{2} A_{1} A_{0} H_{b} \beta_{2}}{a}\\
&f_{49}(t)=\frac{\left(\left(\left(A_{0}^{2} \beta_{1}+\frac{1}{2} A_{0}^{2} \beta_{2}+\Mpl^{2}\right) H_{a}-4 H_{b} \left(\beta_{1}-\frac{\beta_{2}}{4}\right) A_{0}^{2}\right) a^{2}-\left(H_{a} \left(\beta_{1}+\frac{\beta_{2}}{2}\right)+H_{b} \beta_{2}\right) A_{1}^{2}\right) b^{2}}{a^{3}}\\
&f_{50}(t)=\frac{b^{2} \left(2 \beta_{1}+\beta_{2}\right) A_{0}^{2}}{a}\\
&f_{51}(t)=-\frac{b^{2} A_{1}^{2} \left(2 \beta_{1}+\beta_{2}\right) H_{b}}{a^{3}}\\
&f_{52}(t)=\frac{b^{2} \left(2 \beta_{1}+\beta_{2}\right) A_{1}^{2}}{2 a^{3}}\\
&f_{53}(t)=\frac{A_{1}}{a}\\
&f_{54}(t)=\frac{\left(A_{0}^{2} \beta_{1}+A_{0}^{2} \beta_{2}+\Mpl^{2}\right) a^{2}-A_{1}^{2} \beta_{1}}{2 a^{3}}\\
&f_{55}(t)=\frac{6}{4 b \,a^{2}} \left(\left(\left(\beta_{1}+\frac{\beta_{2}}{2}\right) A_{0}^{2}+\frac{\Mpl^{2}}{3}\right) a^{2}-\frac{\left(\beta_{1}+\frac{\beta_{2}}{2}\right) A_{1}^{2}}{3}\right) b \ddot{a}\\& -2 a \left(\left(\left(-6 \beta_{1}-3 \beta_{2}\right) A_{0}^{2}-2 \Mpl^{2}\right) a^{2}+2 A_{1}^{2} \beta_{1}\right) \ddot{b}-2 ab\left(-3 H_{b}^{2} \beta_{1} A_{1}^{2}-\frac{m^{2} A_{1}^{2}}{2}-\frac{\dot{A}_1^{2}}{2}\right)  \\&-2 ab \left(\left(H_{b} \left(H_{b}-4 H_{a}\right) \beta_{1}+2 H_{b}^{2} \beta_{2}+H_{a} H_{b} \beta_{2}-\frac{m^{2}}{2}\right) A_{0}^{2}+3 \left(H_{b}^{2}-\frac{\Lambda}{3}\right) \Mpl^{2}\right) a^{2}\\
&f_{56}(t)=-2 A_{0}^{2} \beta_{1} a \left(H_{a}+H_{b}\right)\\
&f_{57}(t)=-\frac{4 A_{0} \left(\frac{\ddot{a} \left(\beta_{1}+\frac{\beta_{2}}{2}\right) b}{2}+a \left(\left(\beta_{1}+\frac{\beta_{2}}{2}\right) \ddot{b}+\frac{b \left(\left(4 H_{a} H_{b}+2 H_{b}^{2}\right) \beta_{1}+m^{2}\right)}{4}\right)\right)}{b}\\
&f_{58}(t)=-\frac{a A_{0}^{2}}{2}\\
&f_{59}(t)=\frac{4 \left(\beta_{1}+\frac{\beta_{2}}{4}\right) a \ddot{b}+b \left(2 \ddot{a} \beta_{1}+\left(\left(4 H_{a} H_{b}+2 H_{b}^{2}\right) \beta_{1}+H_{a} H_{b} \beta_{2}+H_{b}^{2} \beta_{2}+m^{2}\right) a \right)}{2 b}\\
&f_{60}(t)=\frac{2 A_{0} A_{1} H_{b} \beta_{2}}{a}\\
&f_{61}(t)=-\frac{\beta_{2} A_{1} A_{0}}{2 a}\\
&f_{62}(t)=-\frac{2 \left(\left(H_{a}+H_{b}\right) \beta_{1}+\frac{H_{a} \beta_{2}}{2}\right) A_{1}^{2}}{a}\\
&f_{63}(t)=\frac{A_{1} \beta_{2} \left(H_{a}-2 H_{b}\right)}{2 a}\\
&f_{64}(t)=\frac{2 A_{0} A_{1} \beta_{1} \left(H_{a}+H_{b}\right)}{a}\\
&f_{65}(t)=\frac{\left(3 A_{0}^{2} \beta_{1}+A_{0}^{2} \beta_{2}+3 \Mpl^{2}\right) a^{2}-3 \left(\beta_{1}+\frac{\beta_{2}}{3}\right) A_{1}^{2}}{4 a^{3}}
 \end{split}
\end{equation*}
\begin{equation*}
    \begin{split}
    &f_{66}(t)=\frac{A_{1} \dot{A}_1}{a}\\
&f_{67}(t)=\frac{A_{1}^{2}}{2 a}\\
&f_{68}(t)=-\frac{\dot{A}_1}{a}\\
&f_{69}(t)=-\frac{2 b^{2} A_{1}^{2} \left(2 \beta_{1}+\beta_{2}\right) H_{b}}{a^{3}}\\
&f_{70}(t)=\frac{2 b^{2} A_{1} \left(2 \beta_{1}+\beta_{2}\right) A_{0} H_{b}}{a^{3}}\\
&f_{71}(t)=\frac{b^{2} \left(2 \beta_{1}+\beta_{2}\right) A_{1}^{2}}{a^{3}}\\
&f_{72}(t)=-\frac{b^{2} A_{1} \left(2 \beta_{1}+\beta_{2}\right) A_{0}}{a^{3}}\\
&f_{73}(t)=\frac{\beta_{2} A_{1} A_{0}}{2 a}\\
&f_{74}(t)=-\frac{A_{1} H_{a} \beta_{2}}{2 a}\\
&f_{75}(t)=\frac{\left(3 A_{0}^{2} \beta_{1}+A_{0}^{2} \beta_{2}+3 \Mpl^{2}\right) a^{2}-3 \left(\beta_{1}+\frac{\beta_{2}}{3}\right) A_{1}^{2}}{4 a^{3}}\\
&f_{76}(t)=\frac{2 A_{1}^{2} \beta_{1}}{a}\\
&f_{77}(t)=\frac{A_{1} \beta_{2}}{a}\\
&f_{78}(t)=-\frac{A_{1} \left(2 \beta_{1}+\beta_{2}\right) A_{0}}{a}\\
&f_{79}(t)=-\frac{\left(2 \beta_{1}+\beta_{2}\right) A_{1}^{2}}{2 a}\\
&f_{80}(t)=\frac{A_{1} \left(2 \beta_{1}+\beta_{2}\right) A_{0}}{2 a}\\
&f_{81}(t)=\frac{\left(\left(\left(-\beta_{1}+\beta_{2}\right) A_{0}^{2}+\Mpl^{2}\right) H_{b}-2 \left(\beta_{1}-\frac{\beta_{2}}{4}\right) A_{0}^{2} H_{a}\right) a^{2}-\beta_{1} A_{1}^{2} H_{b}}{a}\\
&f_{82}(t)=a \left(2 \beta_{1}+\beta_{2}\right) A_{0}^{2}\\
&f_{83}(t)=-A_{0} \beta_{2} a \left(H_{a}+H_{b}\right)\\
&f_{84}(t)=a \left(2 \beta_{1}+\beta_{2}\right) A_{0}^{2}\\
&f_{85}(t)=\frac{2 \left(A_{0}^{2} \beta_{1}+A_{0}^{2} \beta_{2}+\Mpl^{2}\right) \left(H_{a}-H_{b}\right) a^{2}-6 \left(H_{a} \left(\beta_{1}+\frac{\beta_{2}}{2}\right)+\frac{H_{b} \beta_{1}}{3}\right) A_{1}^{2}}{4 a}\\
&f_{86}(t)=-a \beta_{1} A_{0}^{2}\\
&f_{87}(t)=-\frac{A_{0} \beta_{2} a \left(H_{a}-H_{b}\right)}{2}\\
&f_{88}(t)=\frac{\left(-A_{0}^{2} \beta_{1}+\Mpl^{2}\right) a^{2}+A_{1}^{2} \beta_{1}}{2 a}\\
&f_{89}(t)=-\frac{\left(2 \beta_{1}+\beta_{2}\right) A_{1}^{2}}{4 a}\\
&f_{90}(t)=2 \left(\frac{\ddot{a} \left(\beta_{1}+\frac{\beta_{2}}{2}\right) b}{2}+a \left(\left(\beta_{1}+\frac{\beta_{2}}{2}\right) \ddot{b}+\frac{b \left(\left(4 H_{a} H_{b}+2 H_{b}^{2}\right) \beta_{1}+m^{2}\right)}{4}\right)\right) A_{0}^{2} b
 \end{split}
\end{equation*}
\begin{equation*}
    \begin{split}
    f_{91}(t)=&2 A_{0}^{2} \frac{5 b^2}{2 }  \left(\beta_{1}+\frac{\beta_{2}}{2}\right)  \ddot{a}+4 A_{0}^{2} \frac{5 b}{2 } \left(\beta_{1}+\frac{\beta_{2}}{2}\right) a \ddot{b}\\&
    + \left(H_{b} \left(H_{b}+2 H_{a}\right) \beta_{1}-H_{b}^{2} \beta_{2}+\frac{3 m^{2}}{10}-2 H_{a} H_{b} \beta_{2}\right) A_{0}^{2}\frac{5 b^2a}{2 }\\&-4 \left(H_{b}+\frac{H_{a}}{2}\right) \left(\beta_{1}+\frac{\beta_{2}}{2}\right)\dot{A}_0 A_{0}\frac{5 b^2a}{2 }-\frac{3 \left(H_{b}^{2}+2 H_{a} H_{b}-\frac{1}{3} \Lambda \right) \Mpl^{2}}{2}b^2a\\&+ b^2\frac{3 \left(A_{1} H_{b}^{2}+\left(-2 A_{1} H_{a}+4 \dot{A}_1\right) H_{b}-2 A_{1} H_{a}^{2}+2 \dot{A}_1 H_{a}\right) A_{1} \beta_{1}}{2 a}\\&+\frac{ b^2}{2 a}3 A_{1} \dot{A}_1 H_{a} \beta_{2}+\frac{5 b^2}{2 a}\frac{m^{2} A_{1}^{2}}{10}+\frac{5 b^2}{2 a}\frac{3 \dot{A}_1^{2}}{10}-\frac{ b^2}{2 a}3 H_{a}^{2} A_{1}^{2} \beta_{2}\\
f_{92}(t)=&-\frac{b \ddot{b}}{4 a}\left(\left(A_{0}^{2} \beta_{1}+A_{0}^{2} \beta_{2}+\Mpl^{2}\right) a^{2}+3 A_{1}^{2} \beta_{1}\right)-\frac{b^2 a}{8 }\left(A_{0}^{2} H_{b}^{2}+4 A_{0}\dot{A}_0 H_{b}+2\dot{A}_0^{2}\right) \beta_{1}\\&-\frac{b^2 a}{4 } A_{0} \left(\beta_{1}+\frac{\beta_{2}}{2}\right) \ddot{A}_0+\frac{3b^2 }{4 a}\left(\beta_{1}+\frac{\beta_{2}}{2}\right)  A_{1} \ddot{A}_1+12 \left(\beta_{1}+\frac{\beta_{2}}{2}\right) A_{1}^{2} \ddot{a}(-\frac{b^2 }{8 a^{2}})\\& -\frac{b^2 a}{8 }\left(\left(A_{0}^{2} H_{b}^{2}+4 A_{0}\dot{A}_0 H_{b}+\dot{A}_0^{2}\right) \beta_{2}+\Mpl^{2} H_{b}^{2}+\frac{m^{2} A_{0}^{2}}{2}-\Lambda \,\Mpl^{2}\right)\\&-\frac{3b^2 }{8 a}) \left(\left(-6 H_{a}^{2}+8 H_{a} H_{b}+H_{b}^{2}\right) \beta_{1}+\left(-3 H_{a}^{2}+4 H_{a} H_{b}\right) \beta_{2}+\frac{m^{2}}{2}\right) A_{1}^{2}\\&+ \frac{3b^2 }{2 a})\dot{A}_1 \left(\beta_{1}+\frac{\beta_{2}}{2}\right) \left(H_{b}-2 H_{a}\right) A_{1}+\frac{3b^2 }{2 a}) \left(\beta_{1}+\frac{\beta_{2}}{2}+\frac{1}{4}\right) \dot{A}_1^{2}
\\
f_{93}(t)=&12 \left(\frac{\ddot{a} \left(\beta_{1}+\frac{\beta_{2}}{2}\right) b}{2}+\left(\left(\beta_{1}+\frac{\beta_{2}}{2}\right) \ddot{b}+\frac{\left(6 \left(2 H_{a} H_{b}+H_{b}^{2}\right) \beta_{1}+m^{2}\right) b}{12}\right) a \right) A_{0}^{2} b\\
f_{94}(t)=&\frac{9b}{2 a^{2}}\left(\left(A_{0}^{2} \beta_{1}+\frac{1}{2} A_{0}^{2} \beta_{2}+\frac{1}{3} \Mpl^{2}\right) a^{2}+\frac{\left(\beta_{1}+\frac{\beta_{2}}{2}\right) A_{1}^{2}}{3}\right) \ddot{a}+\frac{3b^2}{2 a}\frac{A_{1}^{2} H_{b} \left(H_{b}+2 H_{a}\right) \beta_{1}}{3}\\& +\left(\frac{H_{a} H_{b} \beta_{2}}{3}-\frac{m^{2}}{6}\right) \frac{3b^2}{2 a}A_{1}^{2}-\frac{3b^2}{2 a}\frac{\dot{A}_1^{2}}{6}+ \frac{3b}{2 a}\left(\left(A_{0}^{2} \beta_{2}+2 A_{0}^{2} \beta_{1}+\frac{2}{3} \Mpl^{2}\right) a^{2}+\frac{2 A_{1}^{2} \beta_{1}}{3}\right) \ddot{b}\\&+ \frac{3b^2a}{2 }\left(H_{b} A_{0}^{2} \left(H_{b}+2 H_{a}\right) \beta_{1}+\frac{\Mpl^{2} H_{b}^{2}}{3}+\frac{2 \Mpl^{2} H_{a} H_{b}}{3}+\frac{m^{2} A_{0}^{2}}{6}+\frac{\Lambda \,\Mpl^{2}}{3}\right)  \\
f_{95}(t)=&2 \left(\frac{\ddot{a} \left(\beta_{1}+\frac{\beta_{2}}{2}\right) b}{2}+a \left(\left(\beta_{1}+\frac{\beta_{2}}{2}\right) \ddot{b}+\frac{b \left(\left(4 H_{a} H_{b}+2 H_{b}^{2}\right) \beta_{1}+m^{2}\right)}{4}\right)\right) A_{0}^{2} b
\end{split}
\end{equation*}

\bibliographystyle{utphys}
\bibliography{paperbib}{}

\end{document}